\documentclass[fleqn,usenatbib]{mnras}

\usepackage{newtxtext}
\usepackage{newtxmath}

\usepackage[T1]{fontenc}

\usepackage{chemformula}
\usepackage{graphicx}
\usepackage{soul}
\usepackage{multirow}
\usepackage{threeparttable}
\usepackage{hyperref}
\usepackage{warpcol}
\usepackage{caption}
\usepackage{booktabs}
\usepackage{lscape}
\usepackage{siunitx}
\usepackage[figuresright]{rotating}
\usepackage{placeins}
\usepackage{amsmath}
\usepackage{bm}
\usepackage{longtable} 
\usepackage{caption}
\usepackage[version=4]{mhchem}
\usepackage{adjustbox}
\usepackage{makecell}

\DeclareRobustCommand{\VAN}[3]{#2}
\let\VANthebibliography\thebibliography
\def\thebibliography{\DeclareRobustCommand{\VAN}[3]{##3}\VANthebibliography}

\usepackage{graphicx}	% Including figure files
\usepackage{amsmath}	% Advanced maths commands

\title[Propanethiols: spectra and ISM search]
{Rotational spectroscopy and astronomical search of propane-1-thiol  (\ce{CH3CH2CH2SH}) and propane-2-thiol (\ce{(CH3)2CHSH})}

\author[Wentao Song et al.]{
Wentao Song,$^{1,5}$
Assimo Maris,$^{1,2}$
Víctor M. Rivilla,$^{3}$\thanks{E-mail: vrivilla@cab.inta-csic.es}
Miguel Sanz-Novo,$^{4}$
Izaskun Jimenez-Serra,$^{3}$
\newauthor
Luca Evangelisti,$^{1,2}$
and Sonia Melandri,$^{1,2}$\thanks{E-mail: sonia.melandri@unibo.it}
\\
$^{1}$Department of Chemistry "Giacomo Ciamician", University of Bologna, Via Gobetti, 85, 40129 Bologna, Italy 
\\
$^{2}$Interdepartmental Centre for Industrial Aerospace Research (CIRI Aerospace), University of Bologna, Forlì, 47121, Italy
\\
$^{3}$Centro de Astrobiolog\'ia (CAB), CSIC-INTA, Ctra. de Ajalvir km. 4, Torrej\'on de Ardoz, 28850 Madrid, Spain\\
$^{4}$ Max Planck Institute for Extraterrestrial Physics,  Gießenbachstraße 1 85748 Garching, Bayern, Deutschland\\
$^5$ Present address: Nanjing Institute of Atomic Scale Manufacturing, 211800, Nanjing, China
}

\date{Accepted XXX. Received YYY; in original form ZZZ}

\pubyear{2026}

\begin{document}
%%%%%%%%%%%%%%%%%%%%%%%%%%%%%%%%%%%%%
\label{firstpage}
\pagerange{\pageref{firstpage}--\pageref{lastpage}}
\maketitle
\begin{abstract}
Sulfur-bearing organic molecules provide essential insights into the chemical complexity and evolution of the interstellar medium (ISM). We report the laboratory rotational spectroscopic characterization of propanethiol (\ce{C3H7SH}), a sulfur analogue of propanol and a key target for investigating interstellar sulfur chemistry, using free-jet absorption millimeter-wave (FJ-AMMW) spectroscopy in the 59.6--80~GHz range. For 1PT, three conformers, \textit{Aa}, \textit{Ag}, and \textit{Gg}, were observed, with \textit{Gg} reported for the first time. For 2PT, both the \textit{anti} and \textit{gauche} conformers were identified. We report the first assignments of several monosubstituted isotopologues in natural abundance, including $^{34}$S-\textit{anti}-2PT, $^{13}$C1-\textit{anti}-2PT, $^{13}$C2-\textit{anti}-2PT, and $^{34}$S-\textit{gauche}-2PT. Resolvable tunneling splitting arising from large-amplitude motion of the sulfhydryl (-SH) group was observed for \textit{Ag}-1PT and \textit{gauche}-2PT. Global fits incorporating previous measurements yielded accurate rotational constants, quartic centrifugal distortion constants, and vibrational coupling parameters. These spectroscopic data enabled an astronomical search toward the chemically rich Galactic Center molecular cloud G+0.693-0.027. While no evidence for propanethiol was found, we established rigorous 3$\sigma$ upper limits for their column densities: $N$(\textit{Gg}-1PT) $< 3\times 10^{12}$ cm$^{-2}$ and $N$(\textit{g}-2PT) $< 8\times 10^{12}$ cm$^{-2}$.  Thus, 1PT and 2PT are at least 13 and 5 times less abundant than ethanethiol in this source, respectively. The O/S ratio for the propane-1-ol/propane-1-thiol pair is $\geq29$, consistent with the solar value ($\simeq37$), providing a benchmark for sulfur depletion in the ISM. Ultimately, this work provides the precise spectral data necessary for future successful interstellar detections.
\end{abstract}
%%%%%%%%%%%%%%%%%%%%%%%%%%%%%%%%%%%%%
%Select between one and six entries from the list of approved keywords. Don't make up new ones.
%%%%%%%%%%%%%%%%%%%%%%%%%%%%%%%%%%%%%%%%%%%%%%%%%%
%%%%%%%%%%%%%%%%% BODY OF PAPER %%%%%%%%%%%%%%%%%%
%%%%%%%%%%%%%%%%%%%%%%%%%%%%%%%%%%%%%%
\section{Introduction}
%%%%%%%%%%%%%%%%%%%%%%%%%%%%%%%%%%%%%%
Sulfur (\ce{S}) is the tenth most abundant element in the Universe, with a solar abundance of S/H = $1.32 \times 10^{-5}$ \citep{asplund2009chemical}.
Despite its lower cosmic abundance compared to hydrogen, oxygen, and carbon, its high chemical reactivity enables it to play a crucial role in key astrochemical processes.
Interstellar sulfur participates in complex reaction networks that drive the evolution of molecular clouds, shape the composition of interstellar ices, and influence the formation of prebiotic molecules \citep{goicoechea2021bottlenecks, li2022unraveling, mifsud2021sulfur, santos2024formation}.
A thorough understanding of sulfur chemistry and its distribution is therefore essential for constructing a comprehensive model of interstellar chemical evolution.

To date, sulfur-bearing interstellar species are predominantly inorganic compounds, including hydrogen sulfide (\ce{H2S}), carbon monosulfide (\ce{CS}), sulfur monoxide (\ce{SO}), sulfur dioxide (\ce{SO2}), and carbonyl sulfide (\ce{OCS}). Among these, the ratios \ce{SO2}/\ce{SO}, \ce{SO2}/\ce{H2S}, and \ce{OCS}/\ce{H2S} have been proposed as key chemical clocks for probing the physical and chemical conditions in molecular clouds and star-forming regions \citep{Charnley1997,Hatchell1998,wakelam2011sulfur}.
In recent years, a total of 12 S-bearing complex organic molecules (COMs) have also been identified.
The definition of COMs comprises molecules that contain carbon and have more than 5 atoms.
Among these, methanethiol, also known as methyl mercaptan (\ce{CH3SH}), is the simplest thiol and the sulfur analogue of methanol. It has been detected in various interstellar environments, including the prestellar core L1544  \citep{vastel2018sulphur}, the hot molecular core Sagittarius B2 (Sgr B2) toward the Galactic Center \citep{muller2016exploring}, the solar-type protostar IRAS 16293-2422 \citep{majumdar2016detection}, and the Galactic Center molecular cloud G+0.693-0.027 \citep{rodriguez-almeida2021b}. Furthermore, ethanethiol (ethyl mercaptan, \ce{C2H5SH}), the next higher homologue of methanethiol, was tentatively detected toward Orion KL \citep{kolesnikova2014spectroscopic} and finally confirmed toward G+0.693 \citep{rodriguez-almeida2021b}.  
Taken together, these recent detections of S-bearing COMs, including dimethyl sulfide (\ce{(CH3)2S}) toward G+0.693-0.027 \citep{sanz-novo2025}, S-bearing cyclic hydrocarbons \citep{araki2026detection}, the c-\ce{C3H2S} family member identified in TMC-1 \citep{remijan2025missing}, and the carbon-sulfur chains and related species reported toward TMC-1 \citep{cernicharo2021tmc}, suggest that interstellar sulfur chemistry is considerably richer than previously recognized.
These results further indicate that additional, more complex S-containing organic molecules, including longer-chain species, may still remain undetected in the ISM.
 
Despite these discoveries, the 'sulfur depletion problem' remains a major unresolved challenge in interstellar chemistry. While the abundance of gas-phase atomic and ionic sulfur in diffuse clouds aligns with the total cosmic abundance, the total observed gas-phase sulfur-bearing species in dense molecular clouds fall significantly short of this expected budget.
Several hypotheses have been proposed to explain this discrepancy \citep{savage1996interstellar,jenkins2009unified,konstantopoulou2024dune,gondhalekar1985depletion,jimenez2011sulfur,laas2019modeling,fuente2023gas,vidal2017reservoir}.
One of the explanations is that sulfur may participate in the formation of other complex molecules that remain undetected \citep{smith1991search,shingledecker2020efficient}, 
largely due to the scarcity of laboratory spectral data, which complicates their identification and quantification. 
Furthermore, incomplete rotational spectral databases for certain interstellar \ce{S}-bearing molecules may hinder accurate matching and identification in current astronomical observations. 
 
Thiols, also known as mercaptans, are organosulfur compounds characterized by a sulfhydryl group (-SH). They are sulfur analogues of alcohols, with the oxygen atom in the hydroxyl group (-OH) replaced by sulfur. As common S-bearing organic molecules, thiols have attracted significant interest in astrochemistry due to their potential role in tracing the formation and prevalence of prebiotic molecules in the Universe.
For instance, methanethiol (\ce{CH3SH}) acts as a precursor to amino acids like cysteine and methionine \citep{kiene1999dimethylsulfoniopropionate}. 
In addition, the comparison between the abundances of alcohols and thiols, as e.g. the \ce{CH3OH / CH3SH} ratio, could offer valuable insights into sulfur depletion in molecular clouds. 
A high \ce{CH3OH / CH3SH} ratio may indicate a low rate of sulfur released from dust grains, whereas a low ratio could imply a higher availability of atomic sulfur \citep{rodriguez-almeida2021b}.
Motivated by the detection of \textit{gauche}-ethanethiol in the ISM, \citep{kolesnikova2014spectroscopic, rodriguez-almeida2021b}, propanethiol (\ce{C3H7SH}) seemed a promising candidate for future astronomical observations.
Furthermore, the detection of propane-1-ol (\ce{C3H7OH}, \ce{CH3CH2CH2OH} also known as 1-propanol or \textit{n}-propanol) in both the molecular cloud G+0.693 \citep{jimenez-serra2022} and Sgr B2(N2) \citep{belloche2022interstellar,Zingsheim2022}, along with the identification of its isomer propane-2-ol exclusively in Sgr B2(N2) \citep{belloche2022interstellar}, provides a strong rationale for searching for their sulfur analogues in the ISM.
Given the chemical and behavioral similarities between alcohols and thiols, the existence of propanethiol is theoretically plausible, and its confirmation through precise laboratory spectroscopy is warranted.  

\begin{figure}
    \centering
    \includegraphics[width=1.0\linewidth]{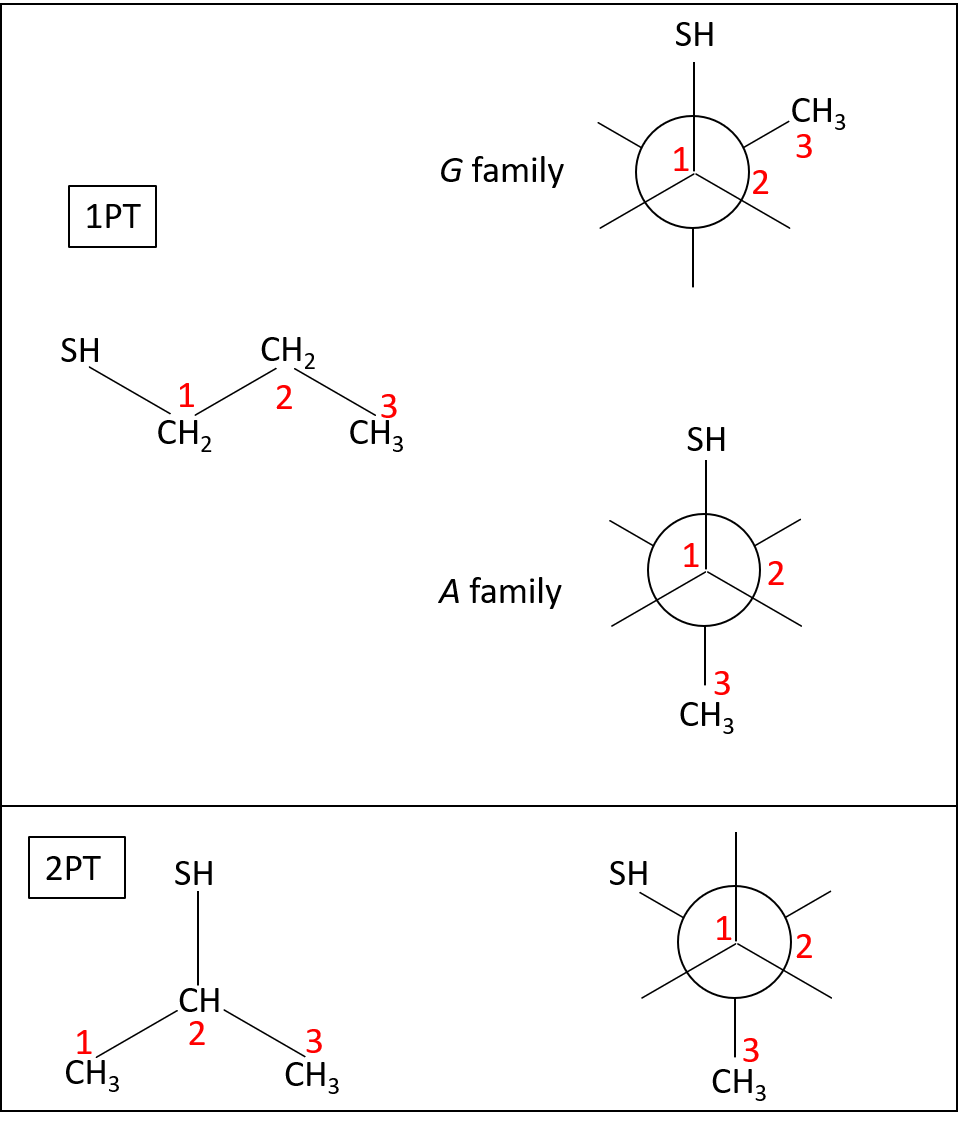}
    \caption{ Sketch and Newman projections of propane-1-thiol (1PT) and propane-2-thiol (2PT) with standard numbering of the carbon atoms.}
    \label{fig:sketch_pt}
\end{figure}

Propanethiol exists in two structural isomers as shown in Figure \ref{fig:sketch_pt}.
In propane-1-thiol (\ce{CH3CH2CH2SH}, from hereon 1PT), also known as \textit{n}-propanethiol or \textit{n}-propylmercaptan or 1-mercaptopropane, the SH group is attached to the terminal carbon atom of the alkyl chain.
In propane-2-thiol (\ce{(CH3)2CHSH}, from hereon 2PT), also known as \textit{iso}-propanethiol or \textit{iso}-propylmercaptan or 2-mercaptopropane, the SH group is attached to the second carbon atom of the alkyl chain.

%%%%%%%%%%%%%%%%%%%%%%%%%%%%%%%%%%%%%%
%%%%%%%%%%%%%%%%%%%%%%%%%%%%%%%%%%%%%%
 
To date, the rotational spectra of both 1PT and 2PT have been investigated in the microwave region. 
As early as 1977, \citet{ohashi1977microwave} investigated the rotational spectrum of 1PT in the 8.5–35 GHz range and observed one conformation.
Subsequently, \citet{nakagawa1981internal} were able to detect an additional conformer exhibiting a tunneling splitting due to a double-minimum potential generated by the rotation of the SH group. 
\citet{griffiths1975microwave} measured the rotational spectrum of 2PT in the 12–40 GHz frequency range, identifying two conformers and observing an analogous splitting in the spectrum of one of them.

In this context, we extended the measurements to %the millimeter wave region
80 GHz and employed quantum mechanical calculations to characterize the conformational space of both molecules.
Using supersonic jet cooling, we can reach low rotational temperatures and high sensitivity.
The determination of highly precise spectroscopic constants including high-order centrifugal distortion constants and spectral information on isotopologues will allow the construction of a rotational spectral catalog as a reliable reference for future astronomical searches for 1PT and 2PT and support further investigations into sulfur chemistry in the ISM. Using the spectroscopic data presented in this work, we have performed a search for 1PT and 2PT toward the G+0.693-0.027 molecular cloud, as discussed before, which represents an excellent astronomical target for molecular detections.
%%%%%%%%%%%%%%%%%%%%%%%%%%%%%%%%%%%%%%
\section{Experimental Methods}
%%%%%%%%%%%%%%%%%%%%%%%%%%%%%%%%%%%%%%
1PT (purity of 99\% and molecular weight of 76.16 g·mol$^{-1}$) and 2PT (purity of 97\%) were purchased from Merck company\footnote{\href{https://www.sigmaaldrich.com}{Merck}} and used without any further purification. 
They are both colorless liquids with a strong, offensive odor.
The boiling points are 340 K and 355 K for 1PT and 2PT, respectively.
The laboratory spectra were recorded within the 59.6-80 GHz range by a Stark-modulated free-jet absorption millimeter-wave (FJ-AMMW) spectrometer, whose details were described previously \citep{Calabrese2013,calabrese2015millimeter,Vigorito2018}.
The estimated uncertainty for the measurements is about 50 kHz, and lines separated by more than 300 kHz are distinguishable.
Argon was used as the carrier gas with a pressure of $P$(Ar)=5 bar and flowed over the sample.
1PT was kept at room temperature, while 2PT was maintained at 258 K.
Then the pressure of the mixture was reduced to $P_0=16.5$ kPa, and expanded into the vacuum chamber through a pinhole nozzle with a diameter of 0.3 mm and a background pressure $P_b\simeq 0.5$ Pa, forming a supersonic jet.
The rotational temperature of the molecules in the jet when using Ar as the carrier gas is estimated to be 5-10 K \citep{Vigorito2018, Sun2025}.
Fitting of the measured transition lines and predictions for the different isomers were performed with Pickett's CALPGM suite of programs \citep{Pickett1991}.
%%%%%%%%%%%%%%%%%%%%%%%%%%%%%%%%%%%%%%
\section{Computational methods}
%%%%%%%%%%%%%%%%%%%%%%%%%%%%%%%%%%%%%%
The theoretical investigation of the structure and dynamics of propanethiols was carried out using the Gaussian 16 software package (revision C.01, Gaussian, Inc., Wallingford, CT, U.S.A.) \citep{g16}.
To explore the conformational space, density functional theory (DFT) was employed using the Becke three-parameter Lee-Yang-Parr hybrid functional (B3LYP~\cite{becke1992density, PhysRevB.37.785}) combined with the D3 version of Grimme's empirical dispersion with Becke-Johnson damping (D3(BJ)~\cite{grimme2011effect}) and the valence triple-$\zeta$ polarized basis set Def2-TZVP~\cite{B508541A}. 
At this level of theory, potential energy surfaces (PESs) were mapped by systematic relaxed scans of the relevant dihedral angles using a step size of $10^{\circ}$. 
The local minima identified from these scans were subsequently subjected to full geometry optimization.
Finally, to obtain a more accurate description of molecular properties, further \textit{ab initio} calculations were performed using second-order M\o{}ller-Plesset perturbation theory (MP2~\cite{moller1934note, head1988mp2}) in conjunction with the Dunning correlation-consistent polarized triple-$\zeta$ basis set augmented with diffuse functions (aug-cc-pVTZ~\cite{dunning1989gaussian}).
For both levels of theory, vibrational frequency calculations within the harmonic approximation were performed on the optimized structures.
This ensured the absence of imaginary frequencies, confirming that the geometries obtained represent true local minima.
%%%%%%%%%%%%%%%%%%%%%%%%%%%%%%%%%%%%%%
\section{Conformational analysis}
%%%%%%%%%%%%%%%%%%%%%%%%%%%%%%%%%%%%%%
The investigated molecules possess rotatable single bonds. This structural flexibility generates a series of stable stereoisomers, defined as conformational isomers or conformers. For organic molecules featuring $sp^3$-hybridized atoms, the lowest-energy, and consequently most stable, spatial configurations correspond to the staggered arrangements of the substituents along the bond.
To describe these specific geometries, and to avoid discrepancies and confusion in the literature, we adopt the IUPAC nomenclature following the recommendation of a recent review on interstellar stereoisomerism \citep{rivilla2026interstellar}. This system parallels traditional nomenclature: staggered arrangements with a dihedral angle of approximately 180° are defined as \textit{antiperiplanar} (\textit{ap}), a term equivalent to the traditional \textit{trans} or \textit{anti}, whereas those with an angle of approximately 60° are termed \textit{synclinal} (\textit{sc}), corresponding to the ordinary \textit{gauche}. Similarly, the -60° (or 300°) orientation is denoted as (\textit{sc})' or \textit{gauche'}. Because the two nomenclatures are essentially synonymous near these angles, for the sake of brevity, we will use the traditional terms \textit{anti} and \textit{gauche} throughout the remainder of this paper.

\begin{figure}
\centering
    \includegraphics[width=1.\columnwidth]{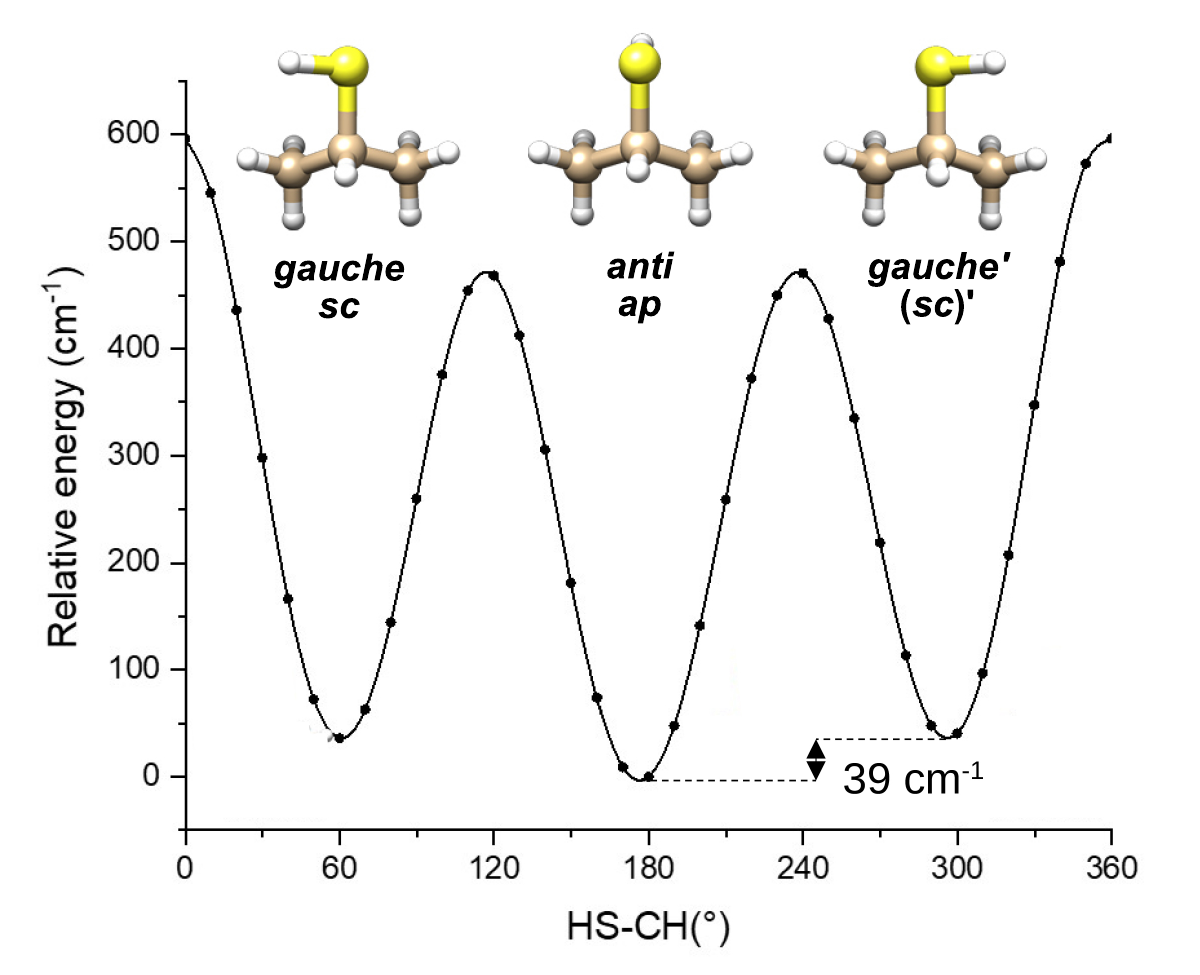} 
\caption{Structure of the 2PT conformers and relaxed potential energy path for the sulfhydryl torsion calculated at the B3LYP-D3(BJ)/Def2-TZVP level. Both the traditional (top) and IUPAC (bottom) nomenclatures are shown.}
\label{fig:iso-PES} 
\end{figure}

Applying these concepts to 2PT, the molecule is characterized by three internal rotation motions associated with the thiol group and the two methyl groups. However, owing to the $C_{3v}$ symmetry of the methyl groups, only rotation around the C–S bond generates non-equivalent species, which are distinguished by the value of their dihedral angle ($\tau_\text{HC-SH}$). The stable conformers of 2PT are therefore the \textit{anti} species and the enantiomeric pair \textit{gauche} and \textit{gauche'}. The torsional PES, is shown in Figure~\ref{fig:iso-PES}, where we explicitly report both the IUPAC and standard nomenclatures.

\begin{figure}
\centering
    \includegraphics[width=1.\columnwidth]{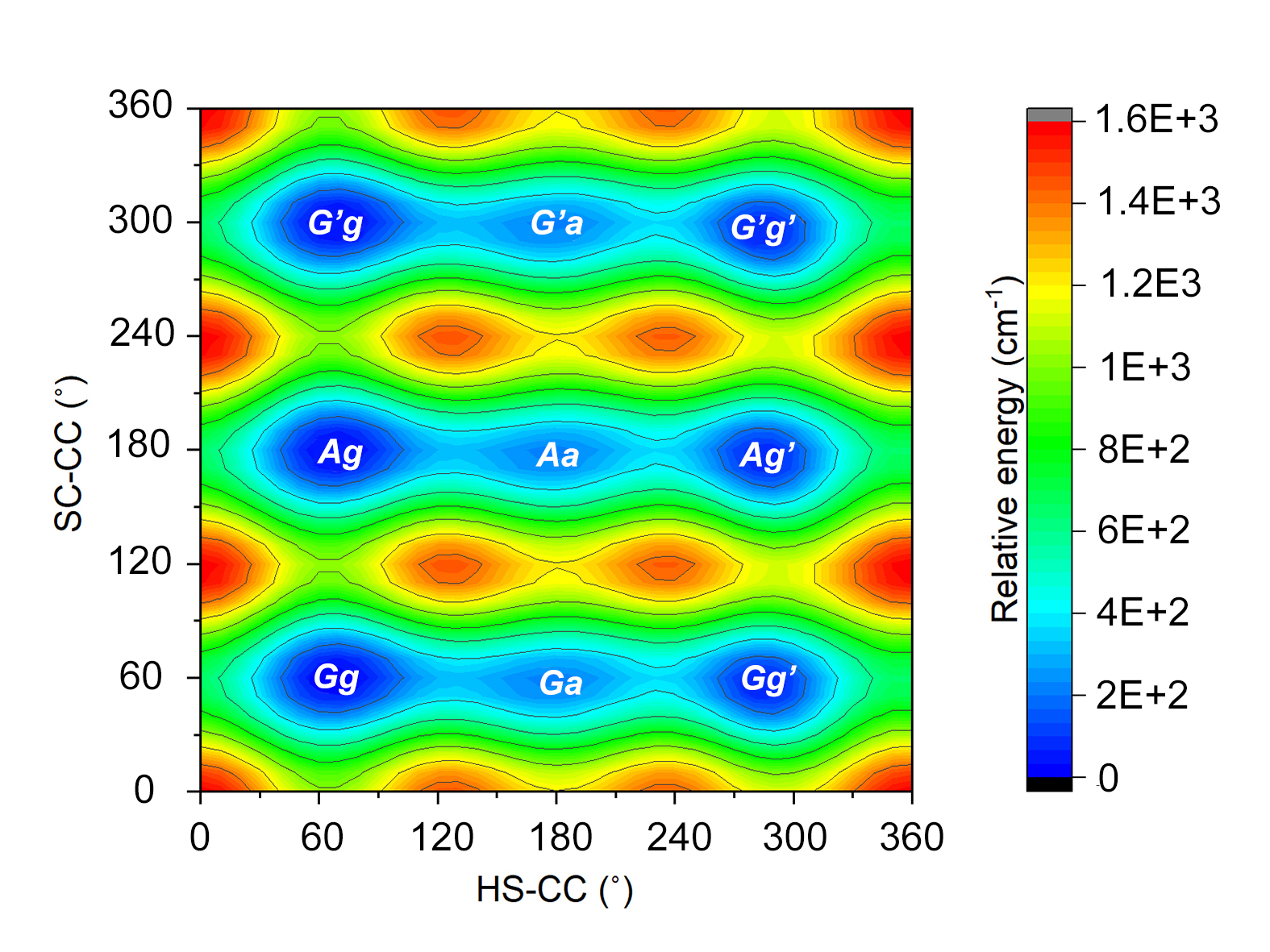} 
\caption{Relaxed bidimensional potential energy surface for the sulfhydryl and skeletal torsions in 1PT calculated at the B3LYP-D3(BJ)/Def2-TZVP level.Blue regions correspond to local minima and identify the possible conformers labelled according to the \textit{anti} or \textit{gauche} orientation of the skeleton chain (capital letter) and the SH group (lower case).}
\label{fig:PES} 
\end{figure}

\begin{figure}
\centering
    \includegraphics[width=1.\columnwidth]{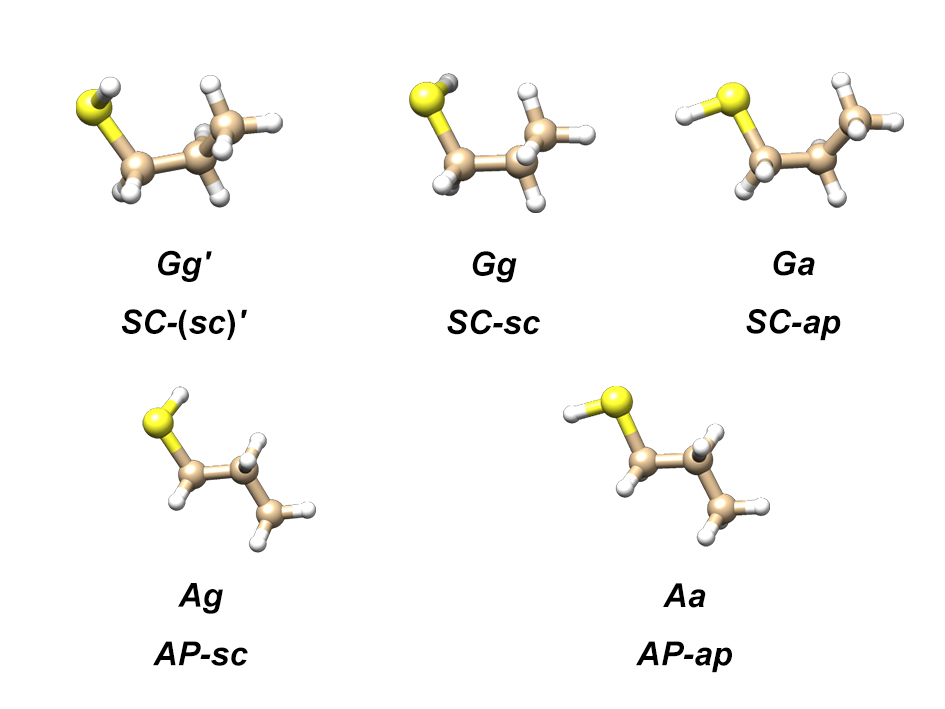} 
\caption{Structure of the five non-equivalent conformers of 1PT. Both the traditional (top) and IUPAC (bottom) nomenclatures are shown.}
\label{fig:1PT Structure} 
\end{figure}

Likewise, 1PT exhibits three torsional motions around single bonds. In this case, however, both the internal rotation of the thiol group ($\tau_\text{CC-SH}$) and the skeletal torsion ($\tau_\text{CC-CS}$) define the conformational PES, while the rotation of the methyl group continues to yield equivalent species. The two-dimensional PES, illustrated in Figure~\ref{fig:PES}, reveals that for each dihedral angle, there are three stable staggered orientations: \textit{gauche}, \textit{anti}, and \textit{gauche'}. The combination of these orientations yields a total of $3^2=9$ possible conformers. These are labeled using a capital letter for the skeletal torsion followed by a lowercase letter for the thiol torsion. With the exception of the all-\textit{anti} species (\textit{Aa}), the remaining conformers exist as enantiomeric pairs: \textit{Ag/Ag'}, \textit{Gg/G'g'}, \textit{Ga/G'a}, and \textit{Gg'/G'g}. Consequently, 1PT possesses five unique, non-equivalent conformers, which are depicted in Figure~\ref{fig:1PT Structure}.

%%%%%%%%%%%%%%%%%%%%%%%%%%%%%%%%%%%%%
%%%%%%%%%%%%%%%%%%%%%%%%%%%%%%%%%%%%%
%%%%%%%%%%%%%%%%%%%%%%%%%%%%%%%%%%%%%
%%%%%%%%%%%%%%%%%%%%%%%%%%%%%%%%%%%%%
\section{Spectral assignment and conformer identification of propane-2-thiol}
%%%%%%%%%%%%%%%%%%%%%%%%%%%%%%%%%%%%%
The DFT and \textit{ab initio} results for the two conformers of 2PT are shown in Table~\ref{tab:isocal}.
According to both theoretical methods, the energy difference between them is very low.
DFT finds the \textit{anti} conformer more stable by about 40 cm$^{-1}$, whereas the two forms are nearly isoenergetic according to \textit{ab initio} calculations. 
Also, the results of the PES scan reported in Figure~\ref{fig:iso-PES} show that the energy barriers between the \textit{anti} and \textit{gauche} conformers are not excessively high (below 600 cm$^{-1}$), so that relaxations and conversions between conformers could occur in the expansion when argon is used as the carrier gas \citep{ruoff1990relaxation}.

%%%%%%%%%%%%%%%%%%%%%%%%%%%%%%%%%%%%%
\begin{table}
    \caption{Calculated spectroscopic parameters for the two conformers of 2PT }
    \centering
    \begin{tabular}{lSSSS}
        \toprule
        &\multicolumn{2}{c}{B3LYP-D3(BJ)/Def2-TZVP}  &\multicolumn{2}{c}{MP2/aug-cc-pVTZ}  \\
        & \textit{anti} & \textit{gauche} & \textit{anti} & \textit{gauche}\\
        \midrule
  $\Delta E_e^a$ (cm$^{-1}$) & 0      & 39     & 3.7    & 0.0     \\
  $\Delta E_0$ (cm$^{-1}$) & 0      & 49     & 0.0    & 5.5     \\
  $\Delta G$ (cm$^{-1}$)   & 0      & 48     & 0.0    & 6.6     \\
  $A$ (MHz)                & 7910.3 & 7901.2 & 7969.8 & 7959.6  \\
  $B$ (MHz)                & 4379.2 & 4489.0 & 4437.4 & 4554.3  \\
  $C$ (MHz)                & 3142.5 & 3152.1 & 3185.5 & 3197.4  \\
  $D_J$ (kHz)              & 1.14    & 1.26    & 1.13    & 1.25     \\
  $D_{JK}$ (kHz)           & 4.21    & 3.95    & 4.36    & 4.07     \\
  $D_K$ (kHz)              & 1.22    & 1.42    & 1.20    & 1.46     \\
  $d_1$ (kHz)              & -0.38   & -0.46   & -0.38   & -0.46    \\
  $d_2$ (kHz)              & -0.11   & -0.14   & -0.11   & -0.14    \\
  $|\mu_a|$ (D)            & 1.62    & 1.42   & 1.67    &  1.45    \\
  $|\mu_b|$ (D)            & 0.00   & 0.62    & 0.00    & 0.61     \\
  $|\mu_c|$ (D)            & 0.31    & 0.64    & 0.28    & 0.64     \\
  $M_{aa}$ (u·Å²)          & 106.17  &  104.48 & 104.56  & 102.77   \\
  $M_{bb}$ (u·Å²)          &  54.65  &  55.86  & 54.08   & 55.29    \\
  $M_{cc}$ (u·Å²)          &   9.23  &  8.11   & 9.33    & 8.20     \\
  $\kappa$                 &  -0.48  & -0.44   & -0.48   & -0.43    \\
        \bottomrule 
    \end{tabular}
    \label{tab:isocal}
    \begin{tabular}{p{0.95\linewidth}}
$^a\Delta E_e$ is the relative electronic energy, $\Delta E_0$ is the relative zero-point-corrected energy, $\Delta G$ is the relative Gibbs free energy (the latter estimated at 298.15 K).
$A$, $B$, and $C$ are the rotational constants and
$D_J, D_{JK}$, $D_K$, $d_1$ and $d_2$ are the quartic centrifugal distortion constants in the S-reduction.
$|\mu_a|$, $|\mu_b|$ and $|\mu_c|$ are the absolute values of the electric dipole moment components. 
$M_{gg}$ ($g$=$a$, $b$ or $c$) are the planar moments of inertia, i.e. $M_{cc} = (I_{aa}+I_{bb}‐I_{cc})/2$.
$\kappa=(2B-A-C)/(A-C)$ is the Ray’s asymmetry parameter.
\end{tabular}
\end{table}
%%%%%%%%%%%%%%%%%%%%%%%%%%%%%%%%%%%%%
\subsection{\textit{Anti}-2PT}
\label{sec:Trans-2PT}
%%%%%%%%%%%%%%%%%%%%%%%%%%%%%%%%%%%%%
The \textit{anti} conformer of 2PT possesses $C_s$ point group symmetry, the $ac$-plane, containing the HC-SH frame, being the symmetry plane.
For the \textit{anti} conformer 60 new transitions of $\mu_a$ and $\mu_c$ R-type with ($J'_{\text{max}}$=12, $K'_{a_{\text{max}}}$=7, $K'_{c_{\text{max}}}$=12) were observed. A global fit including also the 12 lines detected by \citet{griffiths1975microwave} was performed.
Measured transition lines were fitted to Watson's \textit{S}-reduced semirigid asymmetric rotor Hamiltonian \citep{watson1977vibrational} in the $I^r$ representation, including quartic centrifugal distortion constants:
\begin{equation}\begin{aligned}
H &= H_{\text{ROT}} + H_{\text{CD}}
\label{ham}
\end{aligned}\end{equation}
where:
\begin{equation}\begin{aligned}
H_{\text{ROT}} &= A J_a^2 + B J_b^2 + C J_c^2 
\label{hrot}
\end{aligned}\end{equation}
\begin{equation}\begin{aligned}
H_{\text{CD}} &=- D_J J^4 - D_{JK} J^2 (J_a^2 - J_b^2) - D_K J_a^4 
            \\&+ d_{1} J^2 (J_+^2 + J_-^2) + d_{2} (J_+^4 + J_-^4)
\label{hcd}
\end{aligned}\end{equation}
$J_a$, $J_b$ and $J_c$ are the angular momentum principal components.

In addition to the main spectral features, some weak lines could be assigned to the  $^{34}$S and $^{13}$C monosubstituted isotopologues detected in natural abundance. The abundances are 4.2\% and 1.1\%, respectively, for $^{34}$S and $^{13}$C, but it must be noted that due to the symmetry of this conformer, the C1 and C3 position are equivalent; thus, the lines of these monosubstituted isotopologues will appear with an intensity doubled (2.2 \%).
The fitted spectroscopic parameters of all the assigned species are listed in Table \ref{tab:iso} while the measured frequencies are reported in the Appendix: Table ~\ref{tab:2pt-a} and ~\ref{tab:2pt-a-iso}. The new fitting determines, for the parent species, the rotational constants and all four quartic centrifugal distortion constants that were not determined in the previous work by \citet{griffiths1975microwave}.
%%%%%%%%%%%%%%%%%%%%%%%%
\begin{table*}
    \centering
    \caption{Experimental spectroscopic parameters for 2PT .}
%\resizebox{\linewidth}{!}{
    \begin{tabular}{lSSSS} 
        \toprule
        \textit{anti}-2PT& \multicolumn{1}{c}{Normal} & \multicolumn{1}{c}{$^{34}$S} & \multicolumn{1}{c}{$^{13}$C1} & \multicolumn{1}{c}{$^{13}$C2} \\
        \midrule
        $A^a$ (MHz) & 7892.4232(7)$^b$ & 7893.2(2) & 7705.8(6) & 7875.08(5) \\
        $B$ (MHz) & 4414.44079(3) & 4291.525(9) & 4360.36(5) & 4400.268(3) \\
        $C$ (MHz) & 3158.07603(2) & 3094.601(9) & 3100.85(6) & 3153.820(2) \\
        $D_J$ (kHz) & 1.1372(1) & \multicolumn{1}{c}{1.08(3)} & \multicolumn{1}{c}{[1.1372]} & \multicolumn{1}{c}{[1.1372]} \\
        $D_{JK}$ (kHz) & 4.1385(3) & \multicolumn{1}{c}{3.8(2)} & \multicolumn{1}{c}{[4.1385]} & \multicolumn{1}{c}{[4.1385]} \\
        $D_K$ (kHz) & 1.296(2) & \multicolumn{1}{c}{[1.296]$^c$} & \multicolumn{1}{c}{[1.296]} & \multicolumn{1}{c}{[1.296]} \\
        $d_1$ (kHz) & -0.388(7) & \multicolumn{1}{c}{[-0.388]} & \multicolumn{1}{c}{[-0.388]} & \multicolumn{1}{c}{[-0.388]} \\
        $d_2$ (kHz) & -0.108(4) & \multicolumn{1}{c}{[-0.108]} & \multicolumn{1}{c}{[-0.108]} & \multicolumn{1}{c}{[-0.108]} \\
        $N$ & 72$^d$ & 15$^e$ & 6 & 9 \\
        $\sigma$ (kHz) & 41 & 48 & 22 & 26 \\
        Observed type  &\multicolumn{1}{c}{a, c}&\multicolumn{1}{c}{a}&\multicolumn{1}{c}{a}&\multicolumn{1}{c}{a}\\
        $M_{aa}$ (u·Å²)          & 105.239  & 108.520  & 106.650  & 105.460     \\
        $M_{bb}$ (u·Å²)          &  54.789  &  54.789  &  56.331  &  54.783     \\
        $M_{cc}$ (u·Å²)          &   9.245  &   9.242  &   9.253  &   9.391     \\
        \midrule
        \textit{gauche}-2PT& \multicolumn{1}{c}{Normal} & \multicolumn{1}{c}{$^{34}$S} & \\
        \midrule
        $A_0/A_1$ (MHz) & \multicolumn{1}{c}{7877.27700(2) / 7877.42016(2)}  & 7876.19(8) & \\
        $B_0/B_1$ (MHz) & \multicolumn{1}{c}{4524.73521(3) / 4524.76664(3)}  & 4400.70(8) & \\
        $C_0/C_1$ (MHz) & \multicolumn{1}{c}{3172.09792(2) / 3172.10221(2)}  & 3109.98(6) & \\
        $D_J$ (kHz) & 1.2709(9) & \multicolumn{1}{c}{1.24(2)} & \\
        $D_{JK}$ (kHz) & 4.098(4) & \multicolumn{1}{c}{3.7(4)} & \\
        $D_K$ (kHz) & 1.3044(3) & \multicolumn{1}{c}{[1.3044]} & \\
        $d_1$ (kHz) & -0.4619(6) & \multicolumn{1}{c}{[-0.4619]} & \\
        $d_2$ (kHz) & -0.13991(3) & \multicolumn{1}{c}{[-0.13991]} & \\
        $\Delta E_{01}$  (MHz) & 552.2681(1) &552(7)&\\
        $F_{bc}$ (MHz) & 60.5440(2) &52.7(8)&\\
        $F_{ab}$ (MHz) & -96.6534(2) &-99.4(9)& \\
        $N$ & 185$^f$ & 22 &\\
        $\sigma$ (kHz) & 71 & 39 &\\
        Observed type  &\multicolumn{1}{c}{a,b,c}&\multicolumn{1}{c}{a}&\\
        $M_{aa}$ (u·Å²)          & 103.428  &  106.581 &     \\
        $M_{bb}$ (u·Å²)          &  55.892  &  55.885  &      \\
        $M_{cc}$ (u·Å²)          &   8.264  &   8.281  &    \\
        \bottomrule
    \end{tabular}%}
        \vspace{1em}
    \begin{tabular}{p{0.9\linewidth}}
$^a$~$A_v$, $B_v$, and $C_v$ (where $v=0,1$) are the rotational constants of the vibrational sublevels.
$D_J, D_{JK}$, $D_K$, $d_1$ and $d_2$ are the quartic centrifugal distortion constants.
in the S-reduced semirigid rotor Hamiltonian.
$F_{bc}$ and $F_{ab}$ are the Coriolis coupling constants.
$\Delta E_{01}$ is the vibrational energy splitting between the 0 and 1 vibrational states.
$N$ is the number of transition lines in the fit. 
$\sigma$ is the standard deviation of the fit.
$^b$ Error in parentheses is in units of the last digit. 
$^c$ Parameters in the square brackets are fixed to the value of the parent species.
$^d$ including 12 lines from \citet{griffiths1975microwave} 
$^e$ including 7 lines from \citet{griffiths1975microwave} $^f$including 55 lines from \citet{griffiths1975microwave}\\
            \end{tabular}
    \label{tab:iso}
\end{table*}
%%%%%%%%%%%%%%%%%%%%%%%%%%%%%%%%%%%%%
\subsection{\textit{Gauche}-2PT}
\label{sec:Gauche-2PT}
%%%%%%%%%%%%%%%%%%%%%%%%%%%%%%%%%%%%%
The \textit{gauche} conformation shows two equivalent forms, \textit{gauche} and \textit{gauche}' related to the torsion of the thiol group and connected through a relatively low-energy barrier (below 650 cm$^{-1}$ as reported in Figure \ref{fig:iso-PES}). This gives rise to a double-minimum potential energy surface between the two equivalent positions associated with a large-amplitude motion, which splits the ground-state vibrational level into two substates (denoted $v$=0 and $v$=1).
Indeed, because of this motion of the hydrogen atom across the barrier, the rotational transitions show a resolvable tunneling splitting related to transitions within the same or between different vibrational levels depending on symmetry selection rules.
The two vibrational sublevels are associated to wavefunctions of different symmetry with respect to the vibrational coordinate ($v$=0  symmetric and $v$=1 antisymmetric) thus, since the motion involves the inversion of the $\mu_b$ electric dipole moment component, the transitions involving $\mu_a$ and $\mu_c$ components connect rotational levels within the same vibrational state, while transitions involving the $\mu_b$ component connect rotational levels across different vibrational states.
The most prominent feature of the millimeter-wave free jet spectrum is the series of $\mu_a$ R-branch transitions occurring as doublets of almost equal intensity with splittings of about 10 MHz also exhibited by the $\mu_c$, R-branch transitions. 
As described above, for the $\mu_b$ type lines, the interstate transitions are observed. The splitting between the ($0\rightarrow1$) and ($1\rightarrow0$) rotational transitions is around 1100 MHz, which is about twice the energy separation between the levels.
Measured transition lines were fitted to a coupled Hamiltonian based on the reduced axis system expressed as:
\begin{equation}\begin{aligned}
H = H_\text{ROT}^0 + H_\text{ROT}^1 + H_\text{CD} + H_{01}
\label{coupled}
\end{aligned}\end{equation}
where $H_\text{ROT}^0$ and $H_\text{ROT}^1$ correspond to the rigid rotor Hamiltonian (eq. \ref{hrot}) for states $v$=0 and $v$=1, respectively; $H_\text{CD}$ is the quartic centrifugal distortion Hamiltonian (eq. \ref{hcd}) and $H_{01}$ is the interaction Hamiltonian
\begin{equation}
H_{01} = \Delta E_{01} + F_{ab} (J_a J_b + J_b J_a) + F_{bc} (J_b J_c + J_c J_b) 
\end{equation}
Here $\Delta E$ represents the energy difference between the $v$=0 and $v$=1 states, and $F_{ab}$ and $F_{bc}$ are the Coriolis coupling parameters.

Besides the main features of the spectrum, some weak lines could be assigned to the $^{34}$S mono-substituted isotopologue observed in natural abundance.
The fitted spectroscopic parameters for all species are compared in Table~\ref{tab:iso} while the measured transition frequencies are reported in Table~\ref{tab:2pt-g} and ~\ref{tab:2pt-g-iso}.
The new measurements allowed for a significantly improved fitting model with a larger set of parameters with respect to the previous study of \citet{griffiths1975microwave}. In particular, we obtained, for the parent species, distinct rotational constants for each of the two vibrational states (where earlier only an averaged value was reported), as well as all quartic centrifugal distortion constants which were not determined previously.
Notably, the isotopologues of the \textit{gauche} conformer are observed for the first time.
%%%%%%%%%%%%%%%%%%%%%%%%%%%%%%%%%%%%%
\section{Spectral assignment and conformer identification of propane-1-thiol}
%%%%%%%%%%%%%%%%%%%%%%%%%%%%%%%%%%%%%
The theoretical spectroscopic parameters of 1PT are reported in Table~\ref{tab:1ptconst}. As can be seen from the calculated energies, all five conformers lie below 350 cm$^{-1}$ with sizeable values of the dipole moment components. Because some of the energy values are very close (\textit{Gg'}, \textit{Gg} and \textit{Ag}), the energy order of the three lower energy conformers is not univocally determined from the calculations. Depending on the method used and the energy calculated (pure electronic, zero-point corrected or Gibbs free energy) the global minimum is identified either as \textit{Gg'} or \textit{Ag} while \textit{Gg} remains at slightly higher energy (within 60 cm$^{-1}$ from the global minimum).
Previous research by \citet{ohashi1977microwave} and \citet{nakagawa1981internal} reported the presence of \textit{Ag} and \textit{Aa} conformations with a detected tunneling splitting for the \textit{Ag} species. In the current investigation, performed using a supersonic expansion, the \textit{Ag} conformer and one conformer belonging to the \textit{G} family were observed; the latter for the first time. 
%%%%%%%%%%%%%%%%%%%%%%%%%%%%%%%%%%%%%%%%%
\subsection{\textit{Gg}-1PT}
\label{sec:Gg-1PT}
%%%%%%%%%%%%%%%%%%%%%%%%%%%%%%%%%%%%%%%%%
In the experimental spectrum of the newly observed species, 57 transitions ($J'_{\text{max}}$=12, $K'_{a_{\text{max}}}$=6, $K'_{c_{\text{max}}}$=12) involving $\mu_a$, $\mu_b$ and $\mu_c$ R-type lines were detected. All the observed lines are listed in Table A1.
The spectral analysis of the measured transitions reported in Table \ref{tab:1pt-gg} used the same Watson-S reduced Hamiltonian as for \textit{anti}-2PT (eq. \ref{sec:Trans-2PT}) leading to the determination of the rotational constants and the five-quartic centrifugal distortion constants with a root mean square deviation of 20 kHz as reported in Table~\ref{tab:1ptconst}. 

Regarding the conformational assignment, the value of the $A$ rotational constant indicates that the newly assigned species belongs to the \textit{G} family of conformers ($A\simeq$ 12 GHz) rather than to the \textit{T} family ($A\simeq$ 24 GHz).
Based on the theoretical relative energy values (from both B3LYP and MP2 calculations), the least stable conformer, \textit{Ga}, can thus be excluded.
Among the remaining nearly isoenergetic species, \textit{Gg'} and \textit{Gg}, the predicted rotational constants are very similar; however, the two conformers differ significantly in their electric dipole moment components. In particular, the relative intensities of the observed lines transitions are consistent with a $\mu_a$/$\mu_b$ ratio close to 1.
This value agrees well with the prediction for the \textit{Gg} conformer ($\simeq 1$), whereas the corresponding ratio for \textit{Gg'}is substantially larger ($\simeq 4$).
The assignment to the \textit{Gg} species is further supported by comparison of the theoretical and experimental planar moments of inertia, as reported in Table\ref{ham}.
%%%%%%%%%%%%%%%%%%%%%%%%%%%%%%%%%%%%%
\subsection{\textit{Ag}-1PT}
\label{sec:Ag-1PT}
%%%%%%%%%%%%%%%%%%%%%%%%%%%%%%%%%%%%%%%%%
The \textit{Ag} conformer of 1PT exhibits two equivalent structures (\textit{Ag} and \textit{Ag’}), which are connected by a relatively low-energy pathway (below 800 cm$^{-1}$, as shown in Figure \ref{fig:PES}).
In the same way described for the \textit{gauche} conformations of 2PT,  tunneling between the two equivalent \textit{gauche} minima of the thiol internal rotation potential leads to a pair of energy levels, symmetric ($v$=0) and antisymmetric ($v$=1) with respect to the tunneling coordinate. In this case, differently from 2PT, during such a motion, the $\mu_c$ component of the electric dipole moment inverts. 
As a result, the selection rules predict that a- and b-type transitions are allowed for the intrastate 0-0 and 1-1 transitions, whereas c-type transitions arise from interstate 0-1 and 1-0 transitions.
For this conformer, 74 new transitions were identified, including $\mu_a$ and $\mu_c$-type R-branch lines as well as $\mu_c$-type Q-branch lines. 
These new measurements were combined into a global fit together with the 79 transitions reported earlier by \citet{ohashi1977microwave} and \citet{nakagawa1981internal}. Spectral analysis on the measured transitions reported in the Appendix in Table~\ref{tab:1pt-ag} was performed using a coupled Hamiltonian (eq. \ref{coupled}), yielding the spectroscopic parameters listed in Table \ref{tab:1ptconst}. 
%We significantly improved the overall precision of the spectroscopic parameters and determined a larger set of constants than in previous studies. In particular, we obtained distinct rotational constants for each of the two vibrational states (where earlier studies reported only an averaged value), as well as several centrifugal distortion constants ($D_K$ and $d_1$) that had not been determined previously.

We significantly improved the overall precision of the spectroscopic parameters and determined a larger set of constants than in previous studies \citep{nakagawa1981internal}. In particular, we obtained distinct rotational constants for each of the two vibrational states (where earlier only an averaged value was reported), as well as all quartic centrifugal distortion constants (while only two were determined previously). Overall, the new fitting determines thirteen parameters while the previous work determined only seven.

The signals from \textit{Gg'}, \textit{Ga} and \textit{Aa} were not observed in the jet experiment. Since all species posses similar and sizeable dipole moment components' values the lower intensity of their spectra can be attributed to their higher relative energies (conformers \textit{Ga} and \textit{Aa}) or to partial relaxation in the jet. In particular, conformers \textit{Gg'} and \textit{Ga} show low barriers to interconversion around the HS-CC bond (ca. 400 cm$^{-1}$ see Figure~\ref{fig:PES}) and this  could result in relaxation of the higher energy conformers to the lowest one: in this case \textit{Gg}. In a similar way also \textit{Aa} could relax to \textit{Ag} or \textit{Ag'}.A similar relaxation involving the hydroxyl internal rotation has been observed and discussed in the case of 1,2-butanediol \citep{Vigorito2018}.
%%%%%%%%%%%%%%%%%%%%%%%%%%%%%%%%%%%%%
\begin{table*}
    \caption{Calculated and experimental spectroscopic parameters of 1PT.}
    \centering
    \begin{tabular}{l SSSSS}
        \toprule
        \multicolumn{1}{l}{B3LYP-D3(BJ)/Def2-TZVP} & \textit{Gg'} & \textit{Gg}  & \textit{Ga} & \textit{Ag} & \textit{Aa}  \\
        \midrule
        $\Delta E_e$$^a$ (cm$^{-1}$) & 0      & 19     & 323     & 13       & 241      \\
        $\Delta E_0$ (cm$^{-1}$) & 36     & 25     & 320     & 0        & 229      \\
        $\Delta G_0$ (cm$^{-1}$)   & 76     & 37     & 263     & 0        & 193      \\
        $A$ (MHz)                &11654.8 & 11597.0& 11500.1 & 24195.8  & 24124.3  \\
        $B$ (MHz)                & 3192.6 & 3178.8 & 3291.2  & 2330.8   & 2379.7   \\
        $C$ (MHz)                & 2755.6 & 2780.5 & 2821.4  & 2238.1   & 2256.9   \\
        $D_J$ (kHz)              &1.98     & 1.95    & 2.75     & 0.36      & 0.41      \\
        $D_{JK}$ (kHz)           &-10.45   & -9.44   & -13.90  & -2.00     & -2.57     \\
        $D_K$ (kHz)              &47.01    & 44.40   & 52.56  & 58.92     & 60.54     \\
        $d_1$ (kHz)              &-0.51    & -0.48   & -0.84  & 0.02      & 0.03      \\
        $d_2$ (kHz)              &0.04     & -0.03    & -0.07  & 0.00      & 0.00     \\
   $|\mu_a|/|\mu_b|/|\mu_c|$ (D)&\multicolumn{1}{c}{1.50/0.37/0.59}&\multicolumn{1}{c}{1.17/1.04/0.49}&\multicolumn{1}{c}{0.70/1.35/0.55}&\multicolumn{1}{c}{1.58/0.01/0.63}&\multicolumn{1}{c}{1.20/1.11/0.00}\\
        $M_{aa}$ (u·Å²)          & 149.17  & 148.58  & 144.37   & 210.87    & 207.67    \\ 
        $M_{bb}$ (u·Å²)          & 34.23   & 33.18   & 34.76    & 14.94     & 16.25     \\
        $M_{cc}$ (u·Å²)          &  9.13   & 10.40   & 9.19     & 5.95      & 4.70      \\
        $\kappa$                 & -0.90   & -0.91   & -0.89    & -0.99     & -0.99     \\
        \midrule
        \multicolumn{1}{l}{MP2/aug-cc-pVTZ} & \textit{Gg'} & \textit{Gg}  & \textit{Ga} & \textit{Ag} & \textit{Aa}  \\
        \midrule
        $\Delta E$ (cm$^{-1}$)  & 17       & 33      & 258     & 0        & 191             \\
        $\Delta E_0$ (cm$^{-1}$)& 61       & 47      & 274     & 0        & 195             \\
        $\Delta G$ (cm$^{-1}$)  & 96       & 55      & 232     & 0        & 157             \\
        $A$ (MHz)               & 11601.5  & 11570.6 & 11371.7 & 24131.2  & 24072.4         \\
        $B$ (MHz)               & 3263.8   & 3243.2  & 3401.7  & 2361.6   & 2412.7          \\
        $C$ (MHz)               & 2807.1   & 2831.6  & 2890.5  & 2265.4   & 2286.2          \\
        $D_J$ (kHz)             & 2.05      & 1.95     & 2.82     & 0.37      & 0.41             \\
        $D_{JK}$ (kHz)          & -10.30    & -8.90    & -12.89   & -2.13     & -2.72            \\
        $D_K$ (kHz)             & 44.13     & 40.84    & 45.35    & 56.68     & 58.91            \\
        $d_1$ (kHz)             & -0.53     & -0.48    & -0.86    & 0.03      & 0.03             \\
        $d_2$ (kHz)             & 0.04      & 0.03     & -0.07    & 0.00      & 0.00             \\
   $|\mu_a|/|\mu_b|/|\mu_c|$ (D)&\multicolumn{1}{c}{1.54/0.38/0.58}&\multicolumn{1}{c}{1.21/1.04/0.45}&\multicolumn{1}{c}{0.72/1.34/0.51}&\multicolumn{1}{c}{1.64/0.01/0.61}&\multicolumn{1}{c}{1.23/1.09/0.00}\\
        $M_{aa}$  (u·Å²)         & 145.66    & 145.31   & 139.48   & 208.07    & 204.77           \\
        $M_{bb}$  (u·Å²)         & 34.38     & 33.16    & 35.36    & 15.02     & 16.29            \\
        $M_{cc}$  (u·Å²)         & 9.18      & 10.51    & 9.08     & 5.93      & 4.70             \\
        $\kappa$                 & -0.90     & -0.91    & -0.88    & -0.99     & -0.99            \\
        \midrule
        \multicolumn{1}{l}{Experimental}& \textit{Gg'} & \textit{Gg}  & \textit{Ga} & \textit{Ag} & \textit{Aa}$^b$ \\
        \midrule
        $A_0/A_1$(MHz)            & & 11532.848 (6)$^c$  & & \multicolumn{1}{c}{23907.77(1) / 23907.78(1)}   & 23845.24(22) \\
        $B_0/B_1$ (MHz)            & & 3209.142 (3)       & & \multicolumn{1}{c}{2345.707(4) / 2345.681(4)}    &  2393.469(12) \\
        $C_0/C_1$ (MHz)            & & 2800.617 (3)       & & \multicolumn{1}{c}{2250.1847(9) / 2250.1996(9)}   &  2269.621(16) \\
        $D_J$ (kHz)                & &    2.00 (1)        & & 0.364(2)       &  0.51(14)       \\
        $D_{JK}$ (kHz)             & &   -9.62 (3)        & & -1.93(4)       &  \textendash     \\
        $D_K$ (kHz)                & &   43.8 (3)         & & 55(3)          &  \textendash     \\
        $d_1$ (kHz)                & &   -0.499 (3)       & & -0.023(1)      &  \textendash                \\
        $d_2$ (kHz)                & &   -0.031 (3)       & &                & \textendash               \\
        $\Delta E$ (MHz)       & &                    & & 1612.98(1)     &                  \\
        $F_{bc}$ (MHz)             & &                    & & 1.325(4)       &               \\
        $F_{ab}$ (MHz)             & &                    & & 42(1)          &                 \\
        $N$                        & &    62              & & 153$^d$    &26              \\
        $\sigma$ (kHz)             & &    23              & & 56             &62              \\
        Observed type              & &\multicolumn{1}{c}{a, b, c}& &\multicolumn{1}{c}{a, c}&\multicolumn{1}{c}{a,b}\\
        $M_{aa}$  (u·Å²)           & &   147.056          & & 209.452        & 206.313           \\
        $M_{bb}$  (u·Å²)           & &    33.396          & & 15.142         & 16.358          \\
        $M_{cc}$  (u·Å²)           & &    10.425          & & 5.996          & 4.836          \\
        \bottomrule
    \end{tabular}%}
    \label{tab:1ptconst}
    \begin{tabular}{p{0.9\linewidth}}
$^a\Delta E_e$ is the relative electronic energy, $\Delta E_0$ is the relative zero-point-corrected energy, and $\Delta G$ is the relative Gibbs free energy estimated at 298.15 K).
$A_v$, $B_v$, and $C_v$ (where $v=0,1$) are the rotational constants of the vibrational sublevels.
$D_J, D_{JK}$, $D_K$, $d_1$ and $d_2$ are the quartic centrifugal distortion constants.
in the S-reduced semirigid rotor Hamiltonian.
$|\mu_a|$, $|\mu_b|$ and $|\mu_c|$ are the absolute values of the electric dipole moment components. 
$M_{gg} (g = a, b$ or $c)$ are the planar moments of inertia, i.e. $M_{cc} = (I_{aa}+I_{bb}‐I_{cc})/2$.
$\kappa=(2B-A-C)/(A-C)$ is the Ray’s asymmetry parameter.
$F_{bc}$ and $F_{ab}$ are the Coriolis coupling constants.
$\Delta E$ is the vibrational energy splitting between the 0 and 1 vibrational states.
$N$ is the number of transition lines in the fit. 
$\sigma$ is the standard deviation of the fitting.
$^b$ From \citet{nakagawa1981internal}.
$^c$ In parentheses is the error in units of the last digit. 
$^d$ Including 79 transitions of conformer from \citet{ohashi1977microwave} and \citet{nakagawa1981internal}. 
    \end{tabular}
\end{table*}
%%%%%%%%%%%%%%%%%%%%%%%%%%%%%%%%%%%%%
\section{Rotational Spectral Predictions for Astronomical Searches}
%%%%%%%%%%%%%%%%%%%%%%%%%%%%%%%%%%%%%
Using experimentally derived spectroscopic parameters, together with electric dipole moment components and relative energies calculated at the B3LYP-D3(BJ) / Def2-TZVP level, we simulated the spectra of the 2PT and 1PT conformers using the SPCAT program in Pickett’s CALPGM suite of programs \citep{Pickett1991}. The predicted frequencies and intensities for different conformers at different temperatures are provided as Supplementary Material while the rotational partition function's values are given in the Appendix in Tables \ref{tab:q_2pt} and \ref{tab:q_1pt} for 2PT and 1PT respectively.

The rotational spectra of 2PT confirm the existence of two isoenergetic conformers, namely:  \textit{gauche} and \textit{anti}. 
The existence of two equivalent forms for the \textit{gauche} conformation results in a total population twice that of 2PT-\textit{anti} but since the rotational transitions are split by the presence of two vibrational sublevels, this advantage is offset and the spectral intensities only reflect the rotational populations at different temperatures and the effect of the value of the dipole moment component involved.  
Both species have similar values of the $\mu_a$ component, while the  \textit{gauche} form has a higher value of $\mu_c$ and $\mu_b$  the latter being zero in the  \textit{anti} form.

For 2PT, the \textit{gauche} conformer provides a particularly favourable scenario under cold conditions (7.4~K). 
Below 50~GHz, a substantial number of transitions are predicted with velocity uncertainties better than $\Delta v < 0.1~\mathrm{km\,s^{-1}}$, reaching upper-state energies of $E_{\rm up} \simeq 14.95$~K. 
These transitions directly probe the rotational levels expected to be significantly populated in cold molecular clouds with gas temperatures of $\sim10 K$. This makes them particularly suitable for sources such as the Galactic Center molecular cloud G+0.693-0.027 \citep{rivilla_chemical_2017}, where molecular excitation is characterized by low excitation temperatures ($T_{\rm ex} \approx 5-15$ K) due to sub-thermal excitation conditions caused by the relatively low gas densities ($\sim10^{4}$ cm$^{-3}$) \citep{zeng2018}. Therefore, these transitions represent prime candidates for astronomical searches.
The \textit{anti} conformer also exhibits transitions below 50~GHz with comparable accuracy, extending up to $E_{\rm up} \simeq 12.6$~K. 
Although fewer in number, these transitions remain fully adequate for detection in a source with linewidths of 15--20~$\mathrm{km\,s^{-1}}$, where the achieved frequency precision is comfortably within the astrophysical broadening.

For 1PT, five non-equivalent conformations were found by quantum chemical calculations.  
The analysis identified the \textit{Gg'} (equivalent to \textit{G'g}), \textit{Gg} (equivalent to \textit{G'g'} ) and \textit{Ag} conformers as the most stable, while the \textit{Aa} and the \textit{Ga} (equivalent to \textit{G't}) conformations were predicted at higher energies. 
The three lowest-energy species are calculated to be almost isoenergetic, and the energy order changes depending on the method and corrections used to calculate their energies. 
The experiments allowed the observation of three out of five conformers: \textit{Gg}, \textit{Ag} and \textit{Aa} and the absence of the \textit{Gg'} conformer in the free jet spectra seems to indicate that this conformer relaxes onto another form in the expansion, proving that the global minimum should be  \textit{Gg} or \textit{Ag}.  
All three observed conformations possess a $\mu_a$ dipole moment component greater than  1 D  and two of them (Gg and Aa) also a $\mu_b$ component of similar value, while the \textit{Ag} conformation shows a null $\mu_b$ component and a $\mu_c$   component of about 0.6 D. 
As discussed for 2PT, also for 1PT,  the equivalent conformations bear a population twice as large, therefore favouring the contribution of \textit{Gg} and \textit{Ag} species to the overall rotational spectrum for 1PT. 
Nevertheless, the measured rotational spectra indicate that the rotational transitions of the \textit{Ag} species are split due to a large-amplitude motion resulting in a weaker spectrum.
From the discussion above, it is clear that for 1PT a pronounced conformer dependence of the contribution to the rotational spectrum is found. 

Regarding the frequency ranges, the \textit{Ag} conformer shows a significant set of accurately predicted transitions below 50~GHz with $\Delta v < 0.1~\mathrm{km\,s^{-1}}$, reaching upper-state energies of $E_{\rm up} \simeq 28.8$~K. 
These low-energy transitions are particularly relevant for G$+$0.693$-$0.027, where the excitation temperature is modest and the molecular population is expected to reside primarily in the lowest rotational states. 
The \textit{Gg} conformer provides only a limited number of such transitions, restricted to $E_{\rm up} \lesssim 4.5$~K, implying that its detectability in this source would rely on a small subset of very low-lying transitions. 
For the \textit{Aa} conformer, no transitions below 50~GHz satisfy the $\Delta v < 0.1~\mathrm{km\,s^{-1}}$ requirement. This limitation can be attributed to the fact that the available spectroscopic data for the \textit{Aa} conformer are mainly based on measurements reported in previous studies \citep{nakagawa1981internal}, which have a lower precision for the measurements. 
This suggests that further laboratory measurements would enhance its prospects for detection in low-frequency surveys. 
The transitions measured here can also assist in the identification of Galactic cold clouds with typical linewidths of $\sim0.5~\mathrm{km\,s^{-1}}$ and excitation temperatures of $T_{\rm ex} \sim 10$~K (e.g. \citealt{agundez2023abundance}), where the high precision of the low-frequency predictions becomes even more critical. 
At higher temperatures (200~K), the predictive power of the spectroscopic parameters extends up to 300~GHz with velocity uncertainties below $1~\mathrm{km\,s^{-1}}$ for all conformers of both molecules. 
Under these conditions, transitions of 2PT (\textit{gauche} and \textit{anti}) reach upper-state energies of $E_{\rm up} \simeq 238$~K and $181$~K, respectively, while those of 1PT (\textit{Gg}, \textit{Ag}, and \textit{Aa}) extend up to $E_{\rm up} \simeq 168$~K, $378$~K, and $1.16 \times 10^{3}$~K. 
This level of accuracy is sufficient to enable searches towards hot cores, where typical linewidths are 15--20~$\mathrm{km\,s^{-1}}$ (e.g. \citealt{rivilla2017formation}). 
For transitions whose predicted uncertainties exceed $1~\mathrm{km\,s^{-1}}$, improved frequency values would undoubtedly benefit from additional laboratory measurements in the (sub-)millimetre wavelength range.
%%%%%%%%%%%%%%%%%%%%%%%%%%%%%%%%%%%%%
\section{ASTROCHEMICAL IMPLICATIONS: INTERSTELLAR SEARCH TOWARDS THE G+0.693-0.027 MOLECULAR CLOUD}
%%%%%%%%%%%%%%%%%%%%%%%%%%%%%%%%%%%%%
We have searched for both 1PT and 2PT using the ultra-deep spectral survey towards the G+0.693 molecular cloud, a shocked source located in the Sgr B2 complex in the Galactic Center. This cloud is well known for its chemical richness and has been a prolific hunting ground for new molecular detections in recent years (see e.g., \citealt{rivilla2019b,rivilla2020b,rivilla2021a,rivilla2022a,rivilla2023,rodriguez-almeida2021a,zeng2021,jimenez-serra2022,sanz-novo2023}). Particularly, G+0.693 is an excellent candidate because several S-bearing molecules have been detected for the first time in the ISM, including small tetratomic species such as HOCS$^+$ and HNSO (\citealt{sanz-novo2024a,Sanz-Novo2024b}), thioacids such as thioformic acid (HCOSH), thiols like ethanthiol (\ce{C2H5SH};  \citealt{rodriguez-almeida2021b}), as well as other complex S-bearing species such as dimethyl sulfide (\ce{CH3SCH3}; \citealt{sanz-novo2025}) and a 6-membered S-bearing cyclic hydrocarbon \citep{araki2026detection} the largest S-bearing interstellar molecule known so far. Moreover, several conformers of the O-analogue of 1PT, propane-1-ol (\ce{CH3CH2CH2OH}), were also detected in G+0.693 by \citet{jimenez-serra2022}. 

Our analysis was based on a high-sensitivity, broadband spectral survey obtained toward G+0.693 with the Yebes 40m telescope (Guadalajara, Spain). 
Observational details, including frequency coverage, spectral resolution and beam sizes, are given in \citet{rivilla2023} and \citet{sanz-novo2023}. Given the extended nature of the molecular emission in G+0.693 (e.g., \citealt{brunken_interstellar_2010, Jones2012,Li2020,Santa-Maria2021, Colzi2024}), all intensities were expressed in the $T_{\mathrm{A}}^{\ast}$ scale.

\begin{figure*}
\centerline{\resizebox{0.75\hsize}{!}{\includegraphics[angle=0]{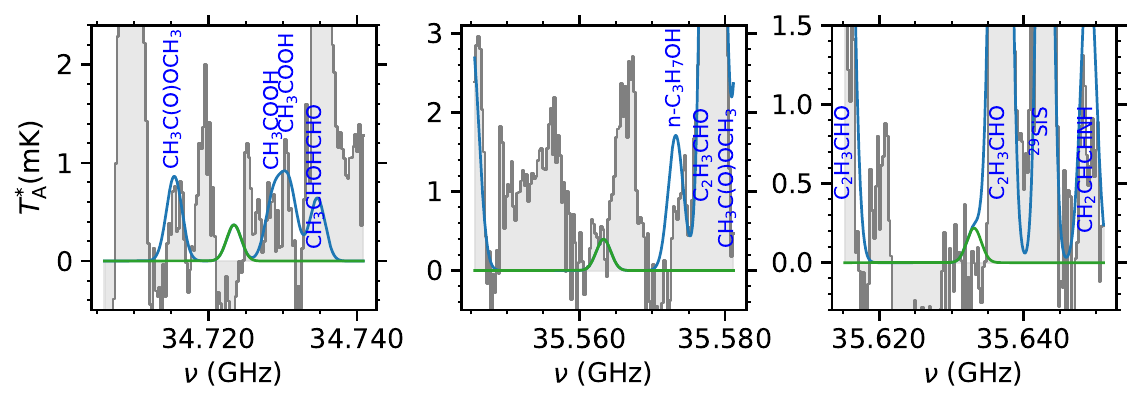}}}
\caption{LTE simulation of the $Gg$-1PT emission using the integrated $3\sigma$ upper limit column density derived toward G+0.693 (in green) together with the cumulative emission model accounting for all the molecular species identified to date in the survey (in blue), both overlaid on the observations (gray histogram).}
\label{f:nondetectionnormal}
\end{figure*}

\begin{figure*}
\centerline{\resizebox{1.0\hsize}{!}{\includegraphics[angle=0]{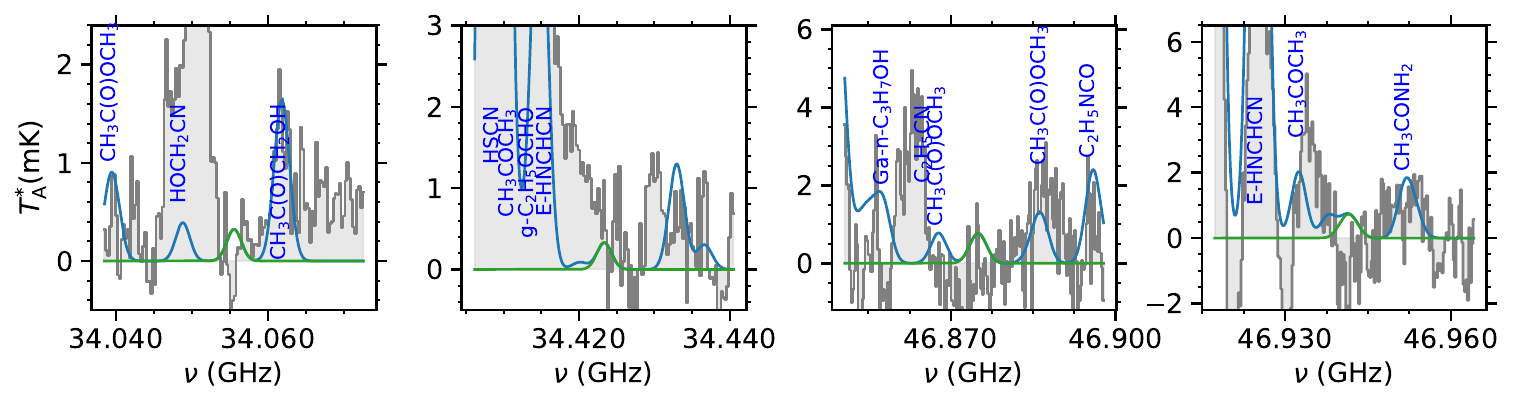}}}
\caption{LTE simulation of the $gauche$-2PT emission using the integrated $3\sigma$ upper limit column density derived toward G+0.693 (in green) together with the cumulative emission model accounting for all the molecular species identified to date in the survey (in blue), both overlaid on the observations (gray histogram).}
\label{f:nondetectioniso}
\end{figure*}

We implemented the newly derived spectroscopic parameters of the most stable conformers of 1PT and 2PT into the MADCUBA package{\footnote{Madrid Data Cube Analysis on ImageJ is a software developed at the Centro de Astrobiología (CAB) in Madrid; \url{http://cab.inta-csic.es/madcuba/}} (\citealt{martin2019}). We generated the Local Thermodynamic Equilibrium (LTE) synthetic spectra using the SLIM (Spectral Line Identification and Modeling) tool of MADCUBA and assuming the excitation temperature ($T_{ex}$), linewidth ($FWHM$) and the local standard of rest velocity ($v_{lsr}$) derived for the structurally related $g$-\ce{C2H5SH} (i.e., $T_{ex} = 10$ K, $FWHM = 20$ km s$^{-1}$ and $v_{lsr}= 69$ km s$^{-1}$ \citealt{rodriguez-almeida2021b}), which were compared to the observational data. 
No signals attributable to either $Gg$-1PT or $gauche$-2PT were found.
To derive an upper limit to the column density, we selected the most intense and least blended transitions predicted by the LTE model, which are shown in Figures \ref{f:nondetectionnormal} and \ref{f:nondetectioniso} (i.e., the $6_{0,6}$--$5_{0,5}$ and  $5_{0,5}$--$4_{0,4}$ transitions, located at 35.563 and 34.423 GHz respectively). Regarding the computation of the upper limit, we computed the line-integrated $3\sigma$ upper limit to the column density ($N$), calculated with MADCUBA-SLIM following the standard procedure described in (\citealt{martin2019}) and represents the maximum $N$ consistent with the non-detection.} This yielded upper limits of $N$($Gg$-1PT) $< 3 \times 10^{12}$ cm$^{-2}$ and $N$($gauche$-2PT) $< 8 \times 10^{12}$ cm$^{-2}$, respectively, corresponding to a molecular abundances relative to \ce{H2} of $< 2 \times 10^{-11}$ and $< 6 \times 10^{-11}$, respectively, adopting $N$(\ce{H2})=$1.35 \times 10^{23}$ cm$^{-2}$ (\citealt{martin_tracing_2008}). 

\begin{figure*}
\centering
    \includegraphics[width=1.75\columnwidth]{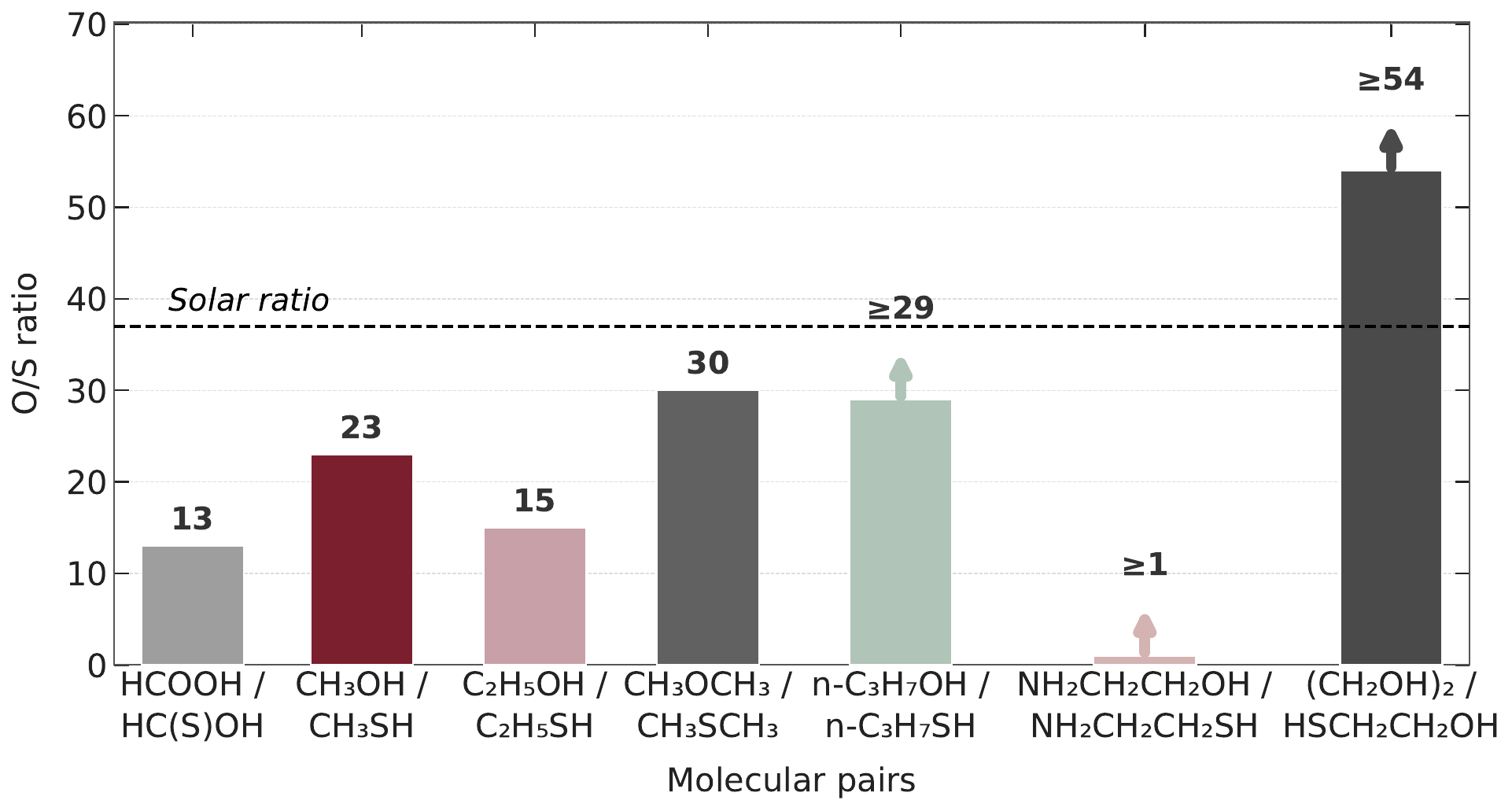} 
\caption{Relative O/S ratio derived toward G+0.693$-$0.027 for all reported S- and O-bearing pairs with $\geq$5 atoms. Lower limits indicated with arrows. Data are taken from: \citet{zeng2018,rodriguez-almeida2021b,rivilla2021a,jimenez-serra2022,Song2022,sanz-novo2025,Bunn_2026} and this work. }
\label{fig:OSratio} 
\end{figure*}

Based on the derived abundances, 1PT and 2PT are $\sim$13 and 5 times less abundant than $g$-\ce{C2H5SH} in G+0.693, which further supports the decrease in abundance expected with increasing molecular complexity already observed for smaller interstellar thiols \citep{rodriguez-almeida2021b}, as well as for other families such as alcohols, aldehydes and isocyanates \citep{jimenez-serra2022,sanz-novo2022,rodriguez-almeida2021a}. In this context, since propane-1-ol has also been detected toward G+0.693, we can also compare the abundance between the this O-bearing and the S-bearing analogue (1PT). Figure \ref{fig:OSratio} shows a comparison of the relative O/S ratio derived for all S- and O-bearing pairs containing more than 5 atoms that have been reported in G+0.693. Interestingly, the propane-1-ol/propane-1-thiol lower limit ratio ($\geq$29, derived from the ratio between the summed abundances of the two detected conformers $Ga$ and $Aa$ of propane-1-ol, and the upper limit derived here for $Gg$-1PT is consistent with the value observed for all other pairs and also close to the solar value (O/S$\sim$37; \citealt{asplund2009chemical}), with the sole exception of the N-bearing species ethanolamine and its S-bearing analogue cysteamine (\ce{NH2CH2CH2SH}; \citealt{Song2022}). Overall, this suggests that the derived upper limit to the abundance might not be far from the true value, although it will need to be confirmed with more sensitive observations.
%%%%%%%%%%%%%%%%%%%%%%%%%%%%%%%%%%%%%
%%%%%%%%%%%%%%%%%%%%%%%%%%%%%%%%%%%%%
%%%%%%%%%%%%%%%%%%%%%%%%%%%%%%%%%%%%%
%%%%%%%%%%%%%%%%%%%%%%%%%%%%%%%%%%%%%
\section{Conclusions}
%%%%%%%%%%%%%%%%%%%%%%%%%%%%%%%%%%%%%
The rotational spectra of 1PT and 2PT were measured in the frequency range 59.6--80 GHz using a FJAMMW spectrometer. For 2PT, two possible conformers, \textit{anti} and \textit{gauche}, were characterized by quantum chemical calculations and subsequently observed in the free-jet expansion. For these species, several monosubstituted isotopologues were also observed in natural abundance: the {$^{34}$S}, {$^{13}$C1}, and {$^{13}$C2} species for the \textit{anti}-2PT conformer, and the {$^{34}$S} species for the \textit{gauche}-2PT conformer.

Exploration of the conformational space of 1PT using quantum chemical methods characterized five non-equivalent conformers, predicting three low-energy forms (\textit{Gg'}, \textit{Gg}, and \textit{Ag}) within a few tens of cm$^{-1}$. Their precise energetic ordering depends on the computational level as well as the inclusion of zero-point and thermal corrections. In this study, the \textit{Ag} and \textit{Gg} conformers were observed and analyzed---the latter for the first time---complementing the \textit{Aa} conformer reported in an earlier work. Global fittings integrating previous measurements and accounting for resolvable tunneling splittings, caused by the large-amplitude motion of the sulfhydryl (-SH) group in both the \textit{Ag}-1PT and \textit{gauche}-2PT conformers, enabled the accurate determination of rotational constants, quartic centrifugal distortion constants, and vibrational coupling parameters.

The recent detection of propane-1-ol in both the molecular cloud G+0.693 \citep{jimenez-serra2022} and Sgr B2(N2) \citep{belloche2022interstellar,Zingsheim2022}, alongside the identification of its isomer propane-2-ol in Sgr B2(N2) \citep{belloche2022interstellar}, prompted a search for 1PT and 2PT toward the G+0.693 molecular cloud using the newly derived spectroscopic parameters. Although neither molecule was detected, rigorous 3$\sigma$ upper limits for their column densities were established at $N$(1PT) $< 3 \times$ 10$^{12}$ cm$^{-2}$ and $N$(2PT) $< 8 \times$ 10$^{12}$ cm$^{-2}$. These results indicate that 1PT and 2PT are approximately 13 and 5 times less abundant than ethanethiol in this source, respectively. 
Furthermore, the derived lower limit for the O/S ratio ($\ge$29) for the propane-1-ol/propane-1-thiol pair is consistent with the solar value ($\sim$37), providing a valuable benchmark for understanding sulfur depletion in the ISM. Higher sensitivity observations are required to discover these prebiotically relevant molecules.
%%%%%%%%%%%%%%%%%%%%%%%%%%%%%%%%%%%%%
\section*{Acknowledgements}
%%%%%%%%%%%%%%%%%%%%%%%%%%%%%%%%%%%%%
SM, LE, AM and WS thank the Italian Ministry of Research (PRIN national project: Astrochemistry beyond the second period elements; CUP J43C21000050001; Grant n. 2020AFB3FX\_003) and the University of Bologna for financial support (RFO). SM and AM acknowledge financial support from the Italian Space Agency, ASI, and the Ministry of University and Research, MUR, through the Space It Up project funded under contract no. 2024-5-E.0 – CUP no. I53D24000060005. 
V.M.R. M.S-N and I.J-S. acknowledges support from the grant PID2022-136814NB-I00 by the Spanish Ministry of Science, Innovation and Universities/State Agency of Research MICIU/AEI/10.13039/501100011033 and by ERDF, UE. V.M.R. also acknowledges support from the grant RYC2020-029387-I funded by MICIU/AEI/10.13039/501100011033 and by "ESF, Investing in your future", from the Consejo Superior de Investigaciones Cient{\'i}ficas (CSIC) and the Centro de Astrobiolog{\'i}a (CAB) through the project 20225AT015 (Proyectos intramurales especiales del CSIC), and  from the grant CNS2023-144464 funded by MICIU/AEI/10.13039/501100011033 and by “European Union NextGenerationEU/PRTR”. M.S-N. acknowledges funding from the Alexander von Humboldt foundation under a Humboldt Research Fellowship.  I.J.-S. acknowledges funding by ERC grant OPENS, GA No. 101125858, funded by the European Union.
%%%%%%%%%%%%%%%%%%%%%%%%%%%%%%%%%%%%%%%%%%%%%%%%%%
\section*{Data Availability}
%%%%%%%%%%%%%%%%%%%%%%%%%%%%%%%%%%%%%
The observational data of the G+0.693-0.027 spectral survey used in this article will be shared on reasonable request to vrivilla@cab.inta-csic.es.
The spectroscopic data, including fitting and prediction files, is offered as supplementary material, and additionally have been shared in \href{https://amsacta.unibo.it/id/eprint/9041/}{AMSActa}.
%The inclusion of a Data Availability Statement is a requirement for articles published in MNRAS. Data Availability Statements provide a standardised format for readers to understand the availability of data underlying the research results described in the article. The statement may refer to original data generated in the course of the study or to third-party data analysed in the article. The statement should describe and provide means of access, where possible, by linking to the data or providing the required accession numbers for the relevant databases or DOIs.
\bibliographystyle{mnras}
\bibliography{bib} % if your bibtex file is called example.bib
%%%%%%%%%%%%%%%%%%%%%%%%%%%%%%%%%%%%%%%%%%%%%%%%%%
%%%%%%%%%%%%%%%%% APPENDICES %%%%%%%%%%%%%%%%%%%%%
\clearpage
%%%%%%%%%%%%%%%%%%%%%%%%%%%%%%%%%%%%%%%%%
\onecolumn
\appendix
\section{ROTATIONAL SPECTROSCOPY DATA and Partition functions }

%\section{ROTATIONAL SPECTROSCOPY DATA}

\setlength{\tabcolsep}{3pt}
\newlength{\linelisttablegap}
\newlength{\linelistcolw}
\newcolumntype{C}{>{\centering\arraybackslash}p{\linelistcolw}}
\setlength{\LTcapwidth}{\textwidth}
\setlength{\LTleft}{0pt}
\setlength{\LTright}{\fill}
\setlength{\linelisttablegap}{2.2em}
\setlength{\linelistcolw}{\dimexpr(\textwidth-\linelisttablegap-16\tabcolsep)/8\relax}

\setlength{\LTleft}{\fill}
\setlength{\LTright}{\fill}
\begin{longtable}{crr|crr}
\caption{Measured rotational transition frequencies and corresponding deviations from the model of $anti$-2PT.}
\label{tab:2pt-a}
\\
\toprule
$J'(K_a',K_c')-J''(K_a'',K_c'')$ & $\nu$ (MHz) & O--C &
$J'(K_a',K_c')-J''(K_a'',K_c'')$ & $\nu$ (MHz) & O--C \\
\midrule
\endfirsthead

\multicolumn{6}{c}{\tablename\ \thetable\ -- Continued from previous page} \\
\toprule
$J'(K_a',K_c')-J''(K_a'',K_c'')$ & $\nu$ (MHz) & O--C &
$J'(K_a',K_c')-J''(K_a'',K_c'')$ & $\nu$ (MHz) & O--C \\
\midrule
\endhead

\midrule
\multicolumn{6}{r}{Continued on next page} \\
\endfoot

\bottomrule
\endlastfoot
2(1,2)	-	1(1,1)	&	13888.64	&	0.01	&	9(7,3)	-	8(7,2)	&	69316.65	&	0.01	\\
2(0,2)	-	1(0,1)	&	14861.54	&	-0.04	&	9(7,2)	-	8(7,1)	&	69316.65	&	0.01	\\
2(1,1)	-	1(1,0)	&	16401.29	&	-0.04	&	9(6,4)	-	8(6,3)	&	69588.80	&	0.02	\\
3(1,3)	-	2(1,2)	&	20671.95	&	-0.01	&	9(6,3)	-	8(6,2)	&	69607.20	&	-0.03	\\
3(0,3)	-	2(0,2)	&	21667.59	&	0.03	&	9(4,6)	-	8(4,5)	&	69812.59	&	-0.01	\\
3(2,2)	-	2(2,1)	&	22717.40	&	0.06	&	9(5,5)	-	8(5,4)	&	69938.55	&	0.00	\\
3(2,1)	-	2(2,0)	&	23767.13	&	-0.06	&	9(5,4)	-	8(5,3)	&	70263.71	&	-0.01	\\
3(1,2)	-	2(1,1)	&	24399.86	&	0.09	&	9(2,7)	-	8(2,6)	&	70438.36	&	0.00	\\
4(1,4)	-	3(1,3)	&	27312.74	&	0.01	&	10(2,9)	-	9(2,8)	&	70444.52	&	0.00	\\
4(0,4)	-	3(0,3)	&	28043.82	&	0.02	&	10(1,9)	-	9(1,8)	&	70635.34	&	-0.01	\\
5(1,5)	-	4(1,4)	&	33827.07	&	-0.03	&	7(2,5)	-	6(1,5)	&	71511.68	&	-0.01	\\
5(0,5)	-	4(0,4)	&	34246.42	&	-0.09	&	11(1,11)	-	10(1,10)	&	71915.70	&	0.01	\\
5(3,3)	-	4(2,3)	&	59862.93	&	-0.01	&	11(0,11)	-	10(0,10)	&	71917.94	&	-0.01	\\
6(2,4)	-	5(1,4)	&	60779.49	&	0.01	&	9(4,5)	-	8(4,4)	&	72256.48	&	-0.01	\\
8(3,6)	-	7(3,5)	&	60889.41	&	0.00	&	9(3,6)	-	8(3,5)	&	73752.61	&	0.02	\\
8(7,2)	-	7(7,1)	&	61477.22	&	0.04	&	6(4,2)	-	5(3,2)	&	73852.01	&	-0.05	\\
8(7,1)	-	7(7,0)	&	61477.22	&	0.04	&	6(4,3)	-	5(3,3)	&	74690.97	&	-0.03	\\
8(6,3)	-	7(6,2)	&	61670.08	&	0.10	&	10(3,8)	-	9(3,7)	&	74770.55	&	0.01	\\
8(6,2)	-	7(6,1)	&	61673.94	&	-0.13	&	5(5,0)	-	4(4,0)	&	74859.43	&	0.13	\\
8(5,4)	-	7(5,3)	&	61958.11	&	0.03	&	5(5,1)	-	4(4,1)	&	74863.60	&	-0.11	\\
8(5,3)	-	7(5,2)	&	62065.47	&	0.00	&	6(3,3)	-	5(1,4)	&	76179.99	&	0.05	\\
8(4,5)	-	7(4,4)	&	62073.76	&	0.04	&	10(2,8)	-	9(2,7)	&	76530.35	&	0.02	\\
8(4,4)	-	7(4,3)	&	63320.65	&	-0.02	&	11(2,10)	-	10(2,9)	&	76795.59	&	-0.01	\\
8(2,6)	-	7(2,5)	&	63903.86	&	0.01	&	11(1,10)	-	10(1,9)	&	76881.29	&	-0.01	\\
9(2,8)	-	8(2,7)	&	64048.20	&	-0.01	&	10(8,3)	-	9(8,2)	&	76967.17	&	0.01	\\
6(3,3)	-	5(2,3)	&	64414.21	&	0.03	&	10(8,2)	-	9(8,1)	&	76967.17	&	0.01	\\
9(1,8)	-	8(1,7)	&	64448.24	&	0.05	&	10(7,4)	-	9(7,3)	&	77211.56	&	0.08	\\
8(3,5)	-	7(3,4)	&	65448.52	&	0.00	&	10(7,3)	-	9(7,2)	&	77214.18	&	-0.10	\\
10(1,10)	-	9(1,9)	&	65599.58	&	0.01	&	7(2,6)	-	6(1,6)	&	77332.54	&	0.00	\\
10(0,10)	-	9(0,9)	&	65605.52	&	0.02	&	10(4,7)	-	9(4,6)	&	77376.16	&	-0.02	\\
5(4,1)	-	4(3,1)	&	66592.63	&	0.01	&	10(6,5)	-	9(6,4)	&	77574.98	&	0.01	\\
5(4,2)	-	4(3,2)	&	66828.61	&	0.00	&	10(6,4)	-	9(6,3)	&	77640.94	&	-0.01	\\
9(3,7)	-	8(3,6)	&	67940.46	&	-0.02	&	10(5,6)	-	9(5,5)	&	77925.98	&	-0.01	\\
6(3,4)	-	5(2,4)	&	68753.44	&	-0.03	&	12(1,12)	-	11(1,11)	&	78230.83	&	0.00	\\
9(8,2)	-	8(8,1)	&	69139.11	&	-0.01	&	12(0,12)	-	11(0,11)	&	78231.66	&	-0.01	\\
9(8,1)	-	8(8,0)	&	69139.11	&	-0.01	&	10(5,5)	-	9(5,4)	&	78743.35	&	0.01	\\

\end{longtable}
%%%%%%%%%%%%%%%%%%%%%%%%%%%%%%%%%%%%%%%%%
\clearpage
%%%%%%%%%%%%%%%%%%%%%%%%%%%%%%%%%%%%%%%%%
\setlength{\LTleft}{\fill}
\setlength{\LTright}{\fill}
\begin{longtable}{lrrrrrr}
\caption{Measured rotational transition frequencies and corresponding deviations from the model of the $^{34}$S and $^{13}$C mono-substituted isotopologues of $anti$-2PT.}
\label{tab:2pt-a-iso}
\\
\toprule
& \multicolumn{2}{c}{\ce{^{34}S}} & \multicolumn{2}{c}{\ce{^{13}C1}} & \multicolumn{2}{c}{\ce{^{13}C2}} \\
\cmidrule(lr){2-3} \cmidrule(lr){4-5} \cmidrule(lr){6-7}
$J'(K_a',K_c')-J''(K_a'',K_c'')$ & $\nu$ (MHz) & O--C & $\nu$ (MHz) & O--C & $\nu$ (MHz) & O--C \\
\midrule
\endfirsthead

\multicolumn{7}{c}{\tablename\ \thetable\ -- Continued from previous page} \\
\toprule
& \multicolumn{2}{c}{\ce{^{34}S}} & \multicolumn{2}{c}{\ce{^{13}C1}} & \multicolumn{2}{c}{\ce{^{13}C2}} \\
\cmidrule(lr){2-3} \cmidrule(lr){4-5} \cmidrule(lr){6-7}
$J'(K_a',K_c')-J''(K_a'',K_c'')$ & $\nu$ (MHz) & O--C & $\nu$ (MHz) & O--C & $\nu$ (MHz) & O--C \\
\midrule
\endhead

\midrule
\multicolumn{7}{r}{Continued on next page} \\
\endfoot

\bottomrule
\endlastfoot

% ===== data =====
2(0,2)-1(0,1)     & 14520.16 & -0.05 &          &       &          &       \\
2(1,1)-1(1,0)     & 15969.09 & -0.03 &          &       &          &       \\
3(1,3)-2(1,2)     & 20218.93 & -0.16 &          &       &          &       \\
3(0,3)-2(0,2)     & 21216.08 & -0.04 &          &       &          &       \\
3(2,2)-2(2,1)     & 22158.13 & -0.10 &          &       &          &       \\
3(2,1)-2(2,0)     & 23100.27 & -0.13 &          &       &          &       \\
3(1,2)-2(1,1)     & 23775.71 &  0.09 &          &       &          &       \\
8(3,6)-7(3,5)     &          &       & 59956.23 & -0.01 & 60746.60 &  0.03 \\
8(5,4)-7(5,3)     & 60316.86 & -0.01 & 61116.68 &  0.01 & 61792.17 & -0.01 \\
8(5,3)-7(5,2)     & 60394.75 &  0.01 &          &       &          &       \\
8(4,5)-7(4,4)     &          &       & 61210.35 & -0.02 &          &       \\
10(1,10)-9(1,9)   & 64287.52 &  0.05 &          &       & 65501.45 &  0.05 \\
10(0,10)-9(0,9)   &          &       &          &       & 65507.46 & -0.02 \\
9(3,7)-8(3,6)     &          &       & 66862.77 &  0.04 & 67788.65 & -0.02 \\
9(5,5)-8(5,4)     &          &       & 68993.66 &  0.01 & 69749.78 & -0.01 \\
9(6,3)-8(6,2)     & 67761.22 &  0.08 &          &       &          &       \\
11(1,11)-10(1,10) & 70477.22 &  0.00 &          &       & 71809.04 & -0.01 \\
11(0,11)-10(0,10) & 70480.48 & -0.01 &          &       &          &       \\
10(3,8)-9(3,7)    &          &       & 73547.01 & -0.02 & 74611.69 &  0.03 \\
11(2,10)-10(2,9)  &          &       &          &       & 76662.24 & -0.04 \\
12(1,12)-11(1,11) & 76665.68 &  0.02 &          &       &          &       \\
12(0,12)-11(0,11) & 76666.92 & -0.02 &          &       &          &       \\
\end{longtable}
%%%%%%%%%%%%%%%%%%%%%%%%%%%%%%%%%%%%%%%%%
\clearpage
%%%%%%%%%%%%%%%%%%%%%%%%%%%%%%%%%%%%%%%%%
\begin{longtable}{crrrr|crrrr}
\caption{Measured rotational transition frequencies and corresponding deviations from the model of $gauche$-2PT}.
\label{tab:2pt-g}
\\
\toprule
\multirow{2}{*}{$J'(K_a',K_c')-J''(K_a'',K_c'')$}
    & \multicolumn{2}{c}{0--0} & \multicolumn{2}{c}{1--1}
    & \multirow{2}{*}{$J'(K_a',K_c')-J''(K_a'',K_c'')$}
    & \multicolumn{2}{c}{1--0} & \multicolumn{2}{c}{0--1} \\
\cmidrule(lr){2-3} \cmidrule(lr){4-5} \cmidrule(lr){7-8} \cmidrule(lr){9-10}
    & $\nu$ (MHz) & O--C & $\nu$ (MHz) & O--C
    & & $\nu$ (MHz) & O--C & $\nu$ (MHz) & O--C \\
\midrule
\endfirsthead

\multicolumn{10}{c}{\tablename\ \thetable\ -- Continued from previous page} \\
\toprule
\multirow{2}{*}{$J'(K_a',K_c')-J''(K_a'',K_c'')$}
    & \multicolumn{2}{c}{0--0} & \multicolumn{2}{c}{1--1}
    & \multirow{2}{*}{$J'(K_a',K_c')-J''(K_a'',K_c'')$}
    & \multicolumn{2}{c}{1--0} & \multicolumn{2}{c}{0--1} \\
\cmidrule(lr){2-3} \cmidrule(lr){4-5} \cmidrule(lr){7-8} \cmidrule(lr){9-10}
    & $\nu$ (MHz) & O--C & $\nu$ (MHz) & O--C
    & & $\nu$ (MHz) & O--C & $\nu$ (MHz) & O--C \\
\midrule
\endhead

\midrule
\multicolumn{10}{r}{Continued on next page} \\
\endfoot

\bottomrule
\endlastfoot
2(1,2)	-	1(1,1)	&	14034.75	&	-0.02	&	14032.30	&	0.02	&	3(0,3)	-	2(1,2)	&	20076.86	&	-0.01	&	18967.82	&	0.03	\\
2(0,2)	-	1(0,1)	&	15054.39	&	-0.02	&	15051.70	&	-0.16	&	3(1,3)	-	2(0,2)	&	23752.44	&	0.27	&	22642.78	&	-0.07	\\
2(1,1)	-	1(1,0)	&	16743.07	&	-0.05	&	16742.71	&	0.00	&	4(0,4)	-	3(1,3)	&		&		&	26324.00	&	-0.06	\\
3(1,3)	-	2(1,2)	&	20864.94	&	0.01	&	20859.23	&	0.02	&	4(1,4)	-	3(0,3)	&	29417.12	&	0.13	&	28343.89	&	-0.09	\\
2(1,1)	-	1(0,1)	&	21455.20	&	-0.10	&	21453.18	&	-0.10	&	5(0,5)	-	4(1,4)	&	34308.89	&	-0.19	&	33221.90	&	0.06	\\
3(0,3)	-	2(0,2)	&	21860.70	&	-0.06	&	21854.69	&	-0.10	&	5(1,5)	-	4(0,4)	&	35257.09	&	0.04	&	34169.10	&	-0.02	\\
3(2,2)	-	2(2,1)	&	23085.03	&	0.00	&	23067.66	&	-0.09	&	4(2,3)	-	3(1,2)	&	39338.43	&	-0.38	&	38235.21	&	0.19	\\
3(2,1)	-	2(2,0)	&	24318.03	&	-0.04	&	24302.70	&	-0.05	&				&		&		&		&		\\
2(2,0)	-	1(1,0)	&	27136.11	&	0.00	&	27143.46	&	0.07	&				&		&		&		&		\\
4(1,4)	-	3(1,3)	&	27555.97	&	0.05	&	27525.52	&	-0.06	&				&		&		&		&		\\
4(0,4)	-	3(0,3)	&	28215.39	&	-0.09	&	28184.54	&	0.11	&				&		&		&		&		\\
4(2,3)	-	3(2,2)	&	30514.48	&	0.04	&	30511.54	&	0.08	&				&		&		&		&		\\
4(3,2)	-	3(3,1)	&	31319.14	&	0.06	&	31328.12	&	0.19	&				&		&		&		&		\\
4(3,1)	-	3(3,0)	&	31607.83	&	-0.23	&	31617.67	&	-0.10	&				&		&		&		&		\\
4(1,3)	-	3(1,2)	&	32641.60	&	0.05	&	32638.25	&	0.07	&				&		&		&		&		\\
5(1,5)	-	4(1,4)	&	34009.64	&	0.08	&	34055.66	&	0.12	&				&		&		&		&		\\
5(0,5)	-	4(0,4)	&	34423.54	&	0.19	&	34468.62	&	-0.02	&				&		&		&		&		\\
5(2,4)	-	4(2,3)	&	37733.60	&	0.03	&	37727.92	&	0.06	&				&		&		&		&		\\
5(1,4)	-	4(1,3)	&	39858.13	&	-0.07	&	39852.01	&	-0.06	&				&		&		&		&		\\
6(1,6)	-	5(1,5)	&	40497.69	&	-0.02	&	40492.05	&	0.19	&				&		&		&		&		\\
6(0,6)	-	5(0,5)	&	40658.25	&	0.03	&	40651.95	&	0.03	&				&		&		&		&		\\
5(3,3)	-	4(2,3)	&	60332.49	&	-0.01	&	60319.17	&	-0.01	&	4(4,1)	-	3(3,0)	&	59592.57	&	-0.06	&		&		\\
8(3,6)	-	7(3,5)	&	61758.92	&	0.00	&	61750.58	&	-0.02	&	4(4,0)	-	3(3,1)	&	59650.90	&	0.07	&		&		\\
6(2,4)	-	5(1,4)	&	62113.00	&	-0.01	&	62112.55	&	0.00	&	8(2,7)	-	7(1,6)	&	59768.72	&	-0.03	&		&		\\
8(5,4)	-	7(5,3)	&	63181.12	&	-0.05	&	63183.41	&	-0.08	&	9(0,9)	-	8(1,8)	&	60127.49	&	-0.02	&		&		\\
8(4,5)	-	7(4,4)	&	63233.17	&	0.02	&	63238.25	&	-0.01	&	9(1,9)	-	8(0,8)	&	60147.19	&	-0.02	&		&		\\
8(5,3)	-	7(5,2)	&	63328.18	&	0.02	&	63331.27	&	0.00	&	6(3,4)	-	5(2,3)	&	61948.35	&	0.01	&	60866.15	&	-0.05	\\
9(2,8)	-	8(2,7)	&	64627.97	&	0.00	&	64555.48	&	0.02	&	5(3,2)	-	4(2,3)	&	62211.84	&	0.04	&	61129.28	&	-0.03	\\
8(2,6)	-	7(2,5)	&	64699.41	&	-0.02	&	64689.10	&	-0.03	&	9(1,8)	-	8(2,7)	&	64983.70	&	0.04	&	63793.18	&	0.00	\\
9(1,8)	-	8(1,7)	&	64854.94	&	-0.01	&	64781.32	&	-0.03	&	9(2,8)	-	8(1,7)	&	65617.26	&	0.02	&	64425.65	&	0.00	\\
8(4,4)	-	7(4,3)	&	64886.69	&	0.04	&	64896.95	&	-0.01	&	7(3,5)	-	6(2,4)	&	66413.80	&	-0.03	&	65340.81	&	-0.01	\\
6(3,3)	-	5(2,3)	&	64916.45	&	-0.04	&	64905.12	&	0.02	&	10(0,10)	-	9(1,9)	&	66470.10	&	-0.06	&	65354.92	&	-0.03	\\
6(1,5)	-	5(0,5)	&	65726.57	&	0.02	&	65697.59	&	-0.01	&	10(1,10)	-	9(0,9)	&	66477.04	&	0.01	&	65361.83	&	0.02	\\
10(1,10)	-	9(1,9)	&	65914.54	&	0.01	&	65914.09	&	0.01	&	5(4,2)	-	4(3,1)	&	67193.27	&	-0.02	&	66061.19	&	-0.03	\\
10(0,10)	-	9(0,9)	&	65917.84	&	-0.06	&	65917.36	&	-0.07	&	5(4,1)	-	4(3,2)	&	67592.23	&	0.07	&	66460.83	&	-0.03	\\
5(4,1)	-	4(3,1)	&	66691.69	&	0.01	&	66679.23	&	0.01	&	8(3,6)	-	7(2,5)	&		&		&	69416.35	&	0.02	\\
8(3,5)	-	7(3,4)	&	66953.02	&	0.01	&	66951.31	&	-0.01	&	10(1,9)	-	9(2,8)	&	71432.58	&	0.01	&	70324.22	&	0.02	\\
5(4,2)	-	4(3,2)	&	66974.18	&	0.02	&	66962.49	&	0.03	&	10(2,9)	-	9(1,8)	&	71718.40	&	0.01	&	70609.42	&	0.00	\\
6(2,5)	-	5(1,5)	&	67756.99	&	0.02	&	67722.84	&	0.01	&	11(0,11)	-	10(1,10)	&	72808.92	&	-0.04	&	71694.61	&	-0.04	\\
9(3,7)	-	8(3,6)	&	68799.60	&	-0.01	&	68788.39	&	-0.01	&	11(1,11)	-	10(0,10)	&	72811.39	&	0.09	&	71697.06	&	0.08	\\
6(3,4)	-	5(2,4)	&	69515.25	&	-0.02	&	69506.12	&	0.01	&	6(4,3)	-	5(3,2)	&		&		&	73152.61	&	-0.05	\\
9(7,3)	-	8(7,2)	&		&		&	70625.25	&	0.21	&	9(3,7)	-	8(2,6)	&		&		&	73526.80	&	-0.01	\\
9(7,2)	-	8(7,1)	&		&		&	70625.73	&	-0.15	&	10(2,8)	-	9(3,7)	&		&		&	73544.77	&	0.02	\\
9(6,4)	-	8(6,3)	&	70953.80	&	-0.01	&	70947.80	&	0.02	&	6(4,3)	-	5(3,2)	&	74255.63	&	-0.02	&		&		\\
10(2,9)	-	9(2,8)	&	70965.10	&	-0.01	&	70956.10	&	0.00	&	9(3,7)	-	8(2,6)	&	74563.78	&	-0.03	&		&		\\
9(6,3)	-	8(6,2)	&	70980.93	&	-0.01	&	70975.11	&	0.01	&	5(5,1)	-	4(4,0)	&	75379.90	&	-0.14	&	74252.90	&	-0.11	\\
9(4,6)	-	8(4,5)	&	71046.90	&	-0.08	&	71041.65	&	-0.03	&	5(5,0)	-	4(4,1)	&	75387.47	&	0.06	&	74260.49	&	0.08	\\
9(2,7)	-	8(2,6)	&	71056.85	&	-0.02	&	71042.00	&	-0.04	&	6(4,2)	-	5(3,3)	&	75872.48	&	0.06	&		&		\\
10(1,9)	-	9(1,8)	&	71086.42	&	-0.06	&	71076.84	&	-0.03	&	11(2,10)	-	10(1,9)	&	77949.69	&	0.06	&		&		\\
9(5,5)	-	8(5,4)	&	71335.02	&	-0.04	&		&		&				&		&		&		&		\\
9(5,4)	-	8(5,3)	&		&		&	71809.03	&	0.05	&				&		&		&		&		\\
11(1,11)	-	10(1,10)	&	72252.63	&	0.02	&	72252.17	&	0.00	&				&		&		&		&		\\
11(0,11)	-	10(0,10)	&	72253.74	&	-0.04	&	72253.28	&	-0.05	&				&		&		&		&		\\
7(3,4)	-	6(2,4)	&	73073.50	&	0.00	&	73065.60	&	0.02	&				&		&		&		&		\\
7(2,5)	-	6(1,5)	&	73350.55	&	0.01	&	73353.92	&	0.00	&				&		&		&		&		\\
6(4,2)	-	5(3,2)	&	73961.74	&	0.06	&	73993.15	&	0.02	&				&		&		&		&		\\
5(5,0)	-	4(4,0)	&	74816.99	&	0.15	&	74817.83	&	0.10	&				&		&		&		&		\\
5(5,1)	-	4(4,1)	&	74822.67	&	-0.02	&	74823.51	&	-0.10	&				&		&		&		&		\\
6(4,3)	-	5(3,3)	&	75031.93	&	-0.02	&	75065.72	&	-0.06	&				&		&		&		&		\\
10(10,1)	-	10(9,1)	&		&		&	75387.48	&	0.00	&				&		&		&		&		\\
10(10,0)	-	10(9,2)	&		&		&	75387.48	&	0.00	&				&		&		&		&		\\
10(3,8)	-	9(3,7)	&	75605.94	&	0.00	&	75588.95	&	0.02	&				&		&		&		&		\\
10(2,8)	-	9(2,7)	&	77051.68	&	-0.02	&	77030.83	&	0.01	&				&		&		&		&		\\
6(3,3)	-	5(1,4)	&		&		&	77032.52	&	0.00	&				&		&		&		&		\\
11(2,10)	-	10(2,9)	&	77320.64	&	0.06	&	77317.78	&	0.05	&				&		&		&		&		\\
11(1,10)	-	10(1,9)	&	77371.34	&	-0.01	&	77368.21	&	0.01	&				&		&		&		&		\\
7(1,6)	-	6(0,6)	&	77696.37	&	0.03	&	77655.88	&	0.02	&				&		&		&		&		\\
7(3,5)	-	6(2,5)	&	79251.49	&	0.01	&	79244.82	&	0.02	&				&		&		&		&		\\
\end{longtable}
%%%%%%%%%%%%%%%%%%%%%%%%%%%%%%%%%%%%%%%%%
\setlength{\LTleft}{\fill}
\setlength{\LTright}{\fill}
\begin{longtable}{lrrrrrr}
\caption{Measured rotational transition frequencies and corresponding deviations from the model of the $^{34}$S and $^{13}$C mono-substituted isotopologues of $gauche$-2PT.}
\label{tab:2pt-g-iso}
\\
\toprule
& \multicolumn{4}{c}{$^{34}$S} & \multicolumn{2}{c}{$^{13}$C1} \\
\cmidrule(lr){2-5} \cmidrule(lr){6-7}
& \multicolumn{2}{c}{$v'-v'' = 0$--0}
& \multicolumn{2}{c}{$v'-v'' = 1$--1}
& \multicolumn{2}{c}{$v'-v'' = 0$--0} \\
\cmidrule(lr){2-3} \cmidrule(lr){4-5} \cmidrule(lr){6-7}
$J'(K_a',K_c')-J''(K_a'',K_c'')$ & $\nu$ (MHz) & O--C & $\nu$ (MHz) & O--C & $\nu$ (MHz) & O--C \\
\midrule
\endfirsthead

\multicolumn{7}{c}{\tablename\ \thetable\ -- Continued from previous page} \\
\toprule
& \multicolumn{4}{c}{$^{34}$S} & \multicolumn{2}{c}{$^{13}$C1} \\
\cmidrule(lr){2-5} \cmidrule(lr){6-7}
& \multicolumn{2}{c}{$v'-v'' = 0$--0}
& \multicolumn{2}{c}{$v'-v'' = 1$--1}
& \multicolumn{2}{c}{$v'-v'' = 0$--0} \\
\cmidrule(lr){2-3} \cmidrule(lr){4-5} \cmidrule(lr){6-7}
$J'(K_a',K_c')-J''(K_a'',K_c'')$ & $\nu$ (MHz) & O--C & $\nu$ (MHz) & O--C & $\nu$ (MHz) & O--C \\
\midrule
\endhead

\midrule
\multicolumn{7}{r}{Continued on next page} \\
\endfoot

\bottomrule
\endlastfoot

% ===== data =====
8(3,6)-7(3,5)     & 60355.68 & -0.03 & 60347.85 &  0.07 &          &       \\
8(2,6)-7(2,5)     & 63384.97 &  0.03 & 63375.81 & -0.04 &          &       \\
9(1,8)-8(1,7)     & 63662.62 &  0.01 &          &       &          &       \\
9(2,8)-8(2,7)     &          &       &          &       & 63463.55 & -0.39 \\
10(1,10)-9(1,9)   & 64639.37 & -0.02 & 64638.96 &  0.04 & 64760.89 & -0.17 \\
10(0,10)-9(0,9)   & 64644.15 & -0.04 & 64643.68 & -0.01 & 64763.81 & -0.22 \\
9(3,7)-8(3,6)     & 67307.01 & -0.03 & 67296.69 & -0.01 &          &       \\
10(2,9)-9(2,8)    &          &       &          &       & 69743.08 &  0.64 \\
9(2,7)-8(2,6)     &          &       &          &       & 69804.58 &  0.05 \\
10(1,9)-9(1,8)    & 69752.33 &  0.01 &          &       & 69852.30 & -0.32 \\
11(1,11)-10(1,10) & 70854.72 &  0.00 & 70854.32 &  0.07 & 70987.24 &  0.19 \\
11(0,11)-10(0,10) & 70856.38 & -0.08 & 70855.98 &  0.00 & 70988.25 &  0.16 \\
10(3,8)-9(3,7)    & 74032.82 &  0.02 & 74018.59 & -0.03 &          &       \\
12(1,12)-11(1,11) & 77069.28 &  0.00 & 77068.81 &  0.07 &          &       \\
12(0,12)-11(0,11) & 77069.77 & -0.06 & 77069.28 &  0.00 &          &       \\
\end{longtable}
%%%%%%%%%%%%%%%%%%%%%%%%%%%%%%%%%%%%%%%%%
\clearpage
%%%%%%%%%%%%%%%%%%%%%%%%%%%%%%%%%%%%%%%%%
\setlength{\LTleft}{\fill}
\setlength{\LTright}{\fill}
\begin{longtable}{crr|crr}
\caption{Measured rotational transition frequencies and corresponding deviations from the model of $Gg$-1PT.}
\label{tab:1pt-gg}
\\
\toprule
$J'(K_a',K_c')-J''(K_a'',K_c'')$ & $\nu$ (MHz) & O--C &
$J'(K_a',K_c')-J''(K_a'',K_c'')$ & $\nu$ (MHz) & O--C \\
\midrule
\endfirsthead

\multicolumn{6}{c}{\tablename\ \thetable\ -- Continued from previous page} \\
\toprule
$J'(K_a',K_c')-J''(K_a'',K_c'')$ & $\nu$ (MHz) & O--C &
$J'(K_a',K_c')-J''(K_a'',K_c'')$ & $\nu$ (MHz) & O--C \\
\midrule
\endhead

\midrule
\multicolumn{6}{r}{Continued on next page} \\
\endfoot

\bottomrule
\endlastfoot
10(2,9)  -  9(2,8)  & 59724.33  &  -0.03 &  4(3,1)  -  3(2,2)  & 66716.49  &   0.02 \\
10(6,4)  -  9(6,3)  & 60190.75  &  -0.01 & 11(3,8)  - 10(3,7)  & 66737.93  &   0.01 \\
10(6,5)  -  9(6,4)  & 60190.75  &  -0.01 & 13(1,12) - 12(2,11) & 67178.29  &  -0.02 \\
10(5,5)  -  9(5,4)  & 60220.70  &  -0.02 & 11(1,10) - 10(1,9)  & 67348.64  &  -0.01 \\
10(5,6)  -  9(5,5)  & 60220.70  &  -0.02 & 12(0,12) - 11(1,11) & 67718.82  &  -0.01 \\
10(4,7)  -  9(4,6)  & 60275.50  &   0.01 & 11(2,9)  - 10(2,8)  & 67925.89  &   0.01 \\
10(1,10) -  9(0,9)  & 60282.98  &   0.00 &  8(2,7)  -  7(1,6)  & 67987.40  &  -0.01 \\
10(4,6)  -  9(4,5)  & 60285.84  &   0.00 & 12(1,12) - 11(1,11) & 68912.24  &   0.01 \\
10(3,8)  -  9(3,7)  & 60302.92  &   0.02 & 12(0,12) - 11(0,11) & 69313.04  &   0.00 \\
10(3,7)  -  9(3,6)  & 60550.81  &   0.00 & 12(1,12) - 11(0,11) & 70506.44  &   0.00 \\
10(1,9)  -  9(1,8)  & 61414.13  &  -0.01 &  8(2,6)  -  7(1,6)  & 70848.37  &  -0.04 \\
10(2,8)  -  9(2,7)  & 61615.69  &   0.02 & 12(2,11) - 11(2,10) & 71461.25  &  -0.01 \\
11(0,11) - 10(1,10) & 61659.53  &  -0.01 &  7(2,6)  -  6(1,6)  & 71922.16  &   0.02 \\
11(1,11) - 10(1,10) & 63253.74  &   0.00 & 12(6,6)  - 11(6,5)  & 72261.46  &  -0.04 \\
 7(2,6)  -  6(1,5)  & 63372.41  &  -0.01 & 12(6,7)  - 11(6,6)  & 72261.46  &  -0.04 \\
11(0,11) - 10(0,10) & 63754.85  &   0.00 & 12(5,8)  - 11(5,7)  & 72314.08  &   0.08 \\
 6(2,5)  -  5(1,5)  & 64693.22  &   0.03 & 12(5,7)  - 11(5,6)  & 72315.14  &  -0.03 \\
 8(1,7)  -  7(0,7)  & 64935.86  &  -0.02 & 12(3,10) - 11(3,9)  & 72380.18  &   0.02 \\
 7(2,5)  -  6(1,5)  & 65136.74  &   0.02 &  9(2,8)  -  8(1,7)  & 72436.52  &   0.02 \\
11(1,11) - 10(0,10) & 65349.07  &   0.02 &  5(3,3)  -  4(2,2)  & 72563.70  &   0.00 \\
11(2,10) - 10(2,9)  & 65604.92  &   0.00 &  5(3,2)  -  4(2,2)  & 72572.92  &   0.02 \\
 6(2,4)  -  5(1,5)  & 65692.85  &  -0.02 &  5(3,3)  -  4(2,3)  & 72782.41  &  -0.03 \\
11(6,6)  - 10(6,5)  & 66224.11  &   0.02 &  5(3,2)  -  4(2,3)  & 72791.65  &   0.01 \\
11(5,7)  - 10(5,6)  & 66264.29  &   0.03 &  7(2,5)  -  6(1,6)  & 73686.45  &   0.00 \\
11(5,6)  - 10(5,5)  & 66264.71  &  -0.06 & 12(5,7)  - 12(4,8)  & 76318.90  &   0.02 \\
11(4,8)  - 10(4,7)  & 66334.88  &   0.03 & 12(5,8)  - 12(4,9)  & 76393.77  &  -0.02 \\
11(3,9)  - 10(3,8)  & 66344.88  &   0.01 & 10(2,9)  -  9(1,8)  & 76741.84  &  -0.03 \\
11(4,7)  - 10(4,6)  & 66355.41  &   0.01 &  6(3,4)  -  5(2,3)  & 78390.00  &   0.01 \\
 4(3,2)  -  3(2,1)  & 66640.91  &  -0.05 &  6(3,3)  -  5(2,3)  & 78417.55  &   0.04 \\
 4(3,1)  -  3(2,1)  & 66643.28  &   0.02 &  6(3,4)  -  5(2,4)  & 78896.40  &   0.00 \\
 4(3,2)  -  3(2,2)  & 66714.17  &   0.00 &  6(3,3)  -  5(2,4)  & 78923.90  &  -0.03 \\
\end{longtable}
%%%%%%%%%%%%%%%%%%%%%%%%%%%%%%%%%%%%%%%%%
\clearpage
%%%%%%%%%%%%%%%%%%%%%%%%%%%%%%%%%%%%%%%%%
\begin{longtable}{crrrr|crrrr}
\caption{Measured rotational transition frequencies and corresponding deviations from the model of $Ag$-1PT.}
\label{tab:1pt-ag}
\\
\toprule
\multirow{2}{*}{$J'(K_a',K_c')-J''(K_a'',K_c'')$}
    & \multicolumn{2}{c}{0--0} & \multicolumn{2}{c}{1--1}
    & \multirow{2}{*}{$J'(K_a',K_c')-J''(K_a'',K_c'')$}
    & \multicolumn{2}{c}{1--0} & \multicolumn{2}{c}{0--1} \\
\cmidrule(lr){2-3} \cmidrule(lr){4-5} \cmidrule(lr){7-8} \cmidrule(lr){9-10}
    & $\nu$ (MHz) & O--C & $\nu$ (MHz) & O--C
    & & $\nu$ (MHz) & O--C & $\nu$ (MHz) & O--C \\
\midrule
\endfirsthead

\multicolumn{10}{c}{\tablename\ \thetable\ -- Continued from previous page} \\
\toprule
\multirow{2}{*}{$J'(K_a',K_c')-J''(K_a'',K_c'')$}
    & \multicolumn{2}{c}{0--0} & \multicolumn{2}{c}{1--1}
    & \multirow{2}{*}{$J'(K_a',K_c')-J''(K_a'',K_c'')$}
    & \multicolumn{2}{c}{1--0} & \multicolumn{2}{c}{0--1} \\
\cmidrule(lr){2-3} \cmidrule(lr){4-5} \cmidrule(lr){7-8} \cmidrule(lr){9-10}
    & $\nu$ (MHz) & O--C & $\nu$ (MHz) & O--C
    & & $\nu$ (MHz) & O--C & $\nu$ (MHz) & O--C \\
\midrule
\endhead

\midrule
\multicolumn{10}{r}{Continued on next page} \\
\endfoot

\bottomrule
\endlastfoot
2(0,2) - 1(0,1) & 9191.19 & -0.09 & 9191.19 & -0.09 & 7(0,7) - 6(1,5) & & & 11133.58 & 0.05 \\
2(1,1) - 1(1,0) & 9286.93 & -0.09 & 9286.93 & -0.09 & 8(0,8) - 7(1,6) & 12150.02 & -0.06 & 15373.26 & 0.03 \\
3(1,3) - 2(1,2) & 13644.10 & 0.02 & 13644.10 & 0.02 & 15(1,15) - 15(0,15) & 14842.58 & -0.04 & 18064.59 & 0.02 \\
3(0,3) - 2(0,2) & 13786.14 & 0.03 & 13786.14 & 0.03 & 14(1,14) - 14(0,14) & 15424.99 & 0.00 & 18647.28 & -0.03 \\
3(1,2) - 2(1,1) & 13930.26 & -0.04 & 13930.26 & -0.04 & 13(1,13) - 13(0,13) & 15984.31 & -0.01 & 19206.99 & 0.00 \\
4(1,4) - 3(1,3) & 18191.78 & 0.02 & 18191.78 & 0.02 & 9(0,9) - 8(1,7) & 16332.47 & -0.08 & 19555.37 & -0.02 \\
4(0,4) - 3(0,3) & 18379.95 & -0.01 & 18379.95 & -0.01 & 12(1,12) - 12(0,12) & 16517.43 & 0.00 & 19740.48 & 0.04 \\
4(1,3) - 3(1,2) & 18573.18 & -0.08 & 18573.18 & -0.08 & 11(1,11) - 11(0,11) & 17021.37 & 0.02 & 20244.63 & -0.05 \\
5(1,5) - 4(1,4) & 22738.96 & 0.05 & 22740.84 & 0.03 & 10(1,10) - 10(0,10) & 17493.31 & 0.03 & 20716.92 & 0.03 \\
5(0,5) - 4(0,4) & 22972.53 & 0.02 & 22972.53 & 0.02 & 9(1,9) - 9(0,9) & 17930.65 & 0.02 & 21154.52 & 0.02 \\
5(1,4) - 4(1,3) & 23214.17 & -0.05 & 23215.90 & -0.08 & 8(1,8) - 8(0,8) & 18331.05 & 0.02 & 21555.12 & 0.03 \\
6(1,6) - 5(1,5) & 27285.72 & -0.01 & 27281.64 & 0.00 & 7(1,7) - 7(0,7) & 18692.34 & -0.01 & 21916.49 & 0.02 \\
6(0,6) - 5(0,5) & 27563.50 & 0.06 & 27563.50 & 0.06 & 6(1,6) - 6(0,6) & 19012.68 & 0.02 & 22236.23 & 0.00 \\
6(1,5) - 5(1,4) & 27862.41 & -0.05 & 27858.20 & 0.00 & 5(1,5) - 5(0,5) & 19290.35 & 0.04 & 22518.15 & 0.05 \\
7(1,7) - 6(1,6) & 31832.05 & 0.01 & 31832.78 & 0.03 & 4(1,4) - 4(0,4) & 19523.92 & 0.05 & 22749.84 & -0.02 \\
7(0,7) - 6(0,6) & 32152.44 & 0.01 & 32152.44 & 0.01 & 3(1,3) - 3(0,3) & 19712.13 & 0.01 & 22937.87 & -0.14 \\
7(1,6) - 6(1,5) & 32499.42 & 0.05 & 32499.84 & -0.04 & 2(1,2) - 2(0,2) & 19854.09 & -0.05 & 23079.93 & -0.12 \\
13(2,12) - 12(2,11) & 59724.33 & 0.03 & 59724.33 & 0.03 & 10(0,10) - 9(1,8) & 20455.97 & 0.00 & 23678.32 & -0.02 \\
13(4,10) - 12(4,9) & 59751.48 & 0.02 & 59751.48 & 0.02 & 11(0,11) - 10(1,9) & 24517.99 & 0.01 & 27739.78 & -0.02 \\
13(4,9) - 12(4,8) & 59751.48 & 0.02 & 59751.48 & 0.02 & 12(0,12) - 11(1,10) & 28516.10 & 0.04 & 31737.21 & -0.06 \\
13(3,11) - 12(3,10) & 59756.93 & 0.04 & 59756.93 & 0.04 & 13(0,13) - 12(1,11) & 32447.58 & 0.09 & & \\
13(3,10) - 12(3,9) & 59758.90 & 0.05 & 59758.90 & 0.05 & 3(1,2) - 2(0,2) & 34070.60 & -0.16 & & \\
13(2,11) - 12(2,10) & 59838.26 & -0.01 & 59838.26 & -0.01 & 4(1,3) - 3(0,3) & 38857.87 & -0.03 & 42083.78 & 0.08 \\
13(1,12) - 12(1,11) & 60333.53 & -0.03 & 60333.07 & -0.02 & 5(1,4) - 4(0,4) & 43692.17 & -0.03 & & \\
14(1,14) - 13(1,13) & 63635.39 & -0.01 & 63635.39 & -0.01 & 8(1,7) - 7(0,7) & & & 61730.12 & 0.03 \\
15(0,15) - 14(0,14) & 68757.66 & 0.03 & 68757.36 & -0.01 & 11(2,10) - 11(1,10) & 60091.17 & -0.05 & & \\
15(2,14) - 14(2,13) & 68904.18 & 0.00 & 68904.18 & 0.00 & 10(2,9) - 10(1,9) & 60609.04 & 0.00 & 63831.23 & 0.12 \\
15(5,11) - 14(5,10) & 68941.74 & -0.05 & 68941.74 & -0.05 & 10(1,9) - 9(0,9) & & & 71869.45 & 0.04 \\
15(5,10) - 14(5,9) & 68941.74 & -0.05 & 68941.74 & -0.05 & 9(2,8) - 9(1,8) & 61080.90 & 0.05 & & \\
15(4,12) - 14(4,11) & 68945.90 & 0.00 & 68945.57 & 0.03 & 8(2,7) - 8(1,7) & 61506.24 & -0.10 & & \\
15(4,11) - 14(4,10) & 68945.90 & 0.00 & 68945.57 & 0.03 & 7(2,6) - 7(1,6) & 61885.19 & -0.01 & & \\
15(3,13) - 14(3,12) & 68953.87 & -0.01 & 68953.51 & 0.00 & 6(2,5) - 6(1,5) & 62217.21 & 0.00 & & \\
15(3,12) - 14(3,11) & 68957.88 & -0.02 & 68957.50 & -0.04 & 5(2,4) - 5(1,4) & 62502.28 & 0.13 & & \\
15(2,13) - 14(2,12) & 69078.35 & 0.03 & 69078.35 & 0.03 & 4(2,3) - 4(1,3) & 62739.78 & -0.07 & & \\
15(1,14) - 14(1,13) & 69602.37 & -0.02 & 69601.85 & 0.00 & 3(2,1) - 3(1,3) & 63504.22 & 0.09 & & \\
16(1,16) - 15(1,15) & 72713.35 & -0.01 & 72713.35 & -0.01 & 9(1,8) - 8(0,8) & 63549.39 & 0.07 & & \\
16(0,16) - 15(0,15) & 73315.80 & 0.13 & 73315.80 & 0.13 & 5(2,3) - 5(1,5) & 63942.21 & 0.01 & & \\
16(2,15) - 15(2,14) & 73492.83 & -0.06 & 73492.83 & -0.06 & 2(2,1) - 1(1,1) & 72455.15 & -0.23 & 75681.26 & -0.12 \\
16(3,14) - 15(3,13) & 73552.72 & 0.00 & 73552.34 & 0.01 & 11(1,10) - 10(0,10) & 73802.37 & 0.02 & 77024.01 & 0.01 \\
16(3,13) - 15(3,12) & 73558.26 & -0.03 & 73557.85 & -0.05 & 3(2,1) - 2(1,1) & 76862.20 & 0.18 & & \\
16(2,14) - 15(2,13) & 73703.21 & -0.03 & 73703.21 & -0.03 & 3(2,2) - 2(1,2) & 77146.64 & -0.01 & & \\
16(1,15) - 15(1,14) & 74234.58 & -0.02 & 74234.02 & -0.01 & 12(1,11) - 11(0,11) & 79017.04 & 0.03 & & \\
17(1,17) - 16(1,16) & 77250.59 & -0.01 & 77250.59 & -0.01 & & & & & \\
17(0,17) - 16(0,16) & 77869.25 & 0.03 & 77869.25 & 0.03 & & & & & \\
\end{longtable}
%%%%%%%%%%%%%%%%%%%%%%%%%%%%%%%%%%%%%%%%%
%%%%%%%%%%%%%%%%%%%%%%%%%%%%%%%%%%%%%%%%%
\clearpage
%%%%%%%%%%%%%%%%%%%%%%%%%%%%%%%%%
\setlength{\LTleft}{\fill}
\setlength{\LTright}{\fill}
\begin{longtable}{crr}
\caption{Measured rotational transition frequencies and corresponding deviations from the model of $Aa$-1PT$^a$.}
\label{tab:1pt-aa}
\\
\toprule
$J'(K_a',K_c')-J''(K_a'',K_c'')$ & $\nu$ (MHz) & O--C \\
\midrule
\endfirsthead

\multicolumn{3}{c}{\tablename\ \thetable\ -- Continued from previous page} \\
\toprule
$J'(K_a',K_c')-J''(K_a'',K_c'')$ & $\nu$ (MHz) & O--C \\
\midrule
\endhead

\midrule
\multicolumn{3}{r}{Continued on next page} \\
\endfoot

\bottomrule
\endlastfoot
 2(0,2)  -   1(0,1)  &    9325.73    &   0.11  \\
 2(1,1)  -   1(1,0)  &    9449.92    &  -0.09  \\
 7(0,7)  -   6(1,6)  &   12368.30    &   0.03  \\
 3(0,3)  -   2(0,2)  &   13987.18    &   0.11  \\
 3(1,2)  -   2(1,1)  &   14174.68    &   0.04  \\
 8(0,8)  -   7(1,7)  &   17426.56    &  -0.03  \\
 4(1,3)  -   3(1,2)  &   18898.76    &  -0.08  \\
 2(1,1)  -   2(0,2)  &   21700.01    &   0.01  \\
 3(1,2)  -   3(0,3)  &   21887.52    &  -0.06  \\
 4(1,3)  -   4(0,4)  &   22139.60    &   0.05  \\
 5(1,4)  -   5(0,5)  &   22457.60    &   0.09  \\
 9(0,9)  -   8(1,8)  &   22531.23    &   0.05  \\
 6(1,5)  -   6(0,6)  &   22843.42    &  -0.02  \\
 7(1,6)  -   7(0,7)  &   23299.73    &   0.02  \\
 5(0,5)  -   4(0,4)  &   23304.49    &  -0.01  \\
 5(1,4)  -   4(1,3)  &   23622.43    &  -0.02  \\
 8(1,7)  -   8(0,8)  &   23829.14    &   0.08  \\
 9(1,8)  -   9(0,9)  &   24434.59    &   0.02  \\
10(1,9)  -  10(0,10) &   25119.57    &  -0.11  \\
10(0,10) -   9(1,9)  &   27678.29    &  -0.03  \\
 6(0,6)  -   5(0,5)  &   27959.31    &  -0.08  \\
 6(1,5)  -   5(1,4)  &   28345.37    &   0.05  \\
 2(1,2)  -   1(0,1)  &   30654.06    &  -0.02  \\
 7(0,7)  -   6(0,6)  &   32610.96    &  -0.06  \\
 7(1,6)  -   6(1,5)  &   33067.37    &   0.08  \\
 3(1,3)  -   2(0,2)  &   35131.52    &  -0.04  \\

\end{longtable}
$^a$ All transition frequencies from \citet{ohashi1977microwave} and \citet{nakagawa1981internal}
%%%%%%%%%%%%%%%%%%%%%%%%%%%%%%%%%%%%%%%%%%%%%%%%%%
\clearpage
%%%%%%%%%%%%%%%%%%%%%%%%%%%%%%%%%
\begin{center}
\centering
\captionsetup{hypcap=false}
\captionof{table}{Rotational partition functions ($Q_{\rm rot}$) for the conformers of 2PT at selected temperatures.}
\label{tab:q_2pt}
\setlength{\tabcolsep}{12pt}
\begin{tabular}{rrr}
\toprule
$T$ (K) & \textit{gauche}-2PT & \textit{anti}-2PT \\
\midrule
300.000   & 165183.7 & 83697.7 \\
225.000   & 107236.4 & 54338.1 \\
200.000   & 89855.4  & 45531.6 \\
150.000   & 58345.4  & 29565.9 \\
75.000    & 20622.2  & 10451.2 \\
37.500  & 7292.7   & 3696.6 \\
18.750 & 2580.5   & 1308.5 \\
9.375 & 914.1    & 463.8 \\
7.400   & 641.7    & 325.7 \\
2.000     & 91.4     & 46.6 \\
\bottomrule
\end{tabular}
\end{center}
%%%%%%%%%%%%%%%%%%%%%%%%%%%%%%%%%
\begin{center}
\centering
\captionsetup{hypcap=false}
\captionof{table}{Rotational partition functions ($Q_{\rm rot}$) for the conformers of 1PT at selected temperatures.}
\label{tab:q_1pt}
\setlength{\tabcolsep}{10pt}
\begin{tabular}{rrrr}
\toprule
$T$ (K) & \textit{Gg}-1PT & \textit{Ag}-1PT & \textit{Aa}-1PT \\
\midrule
300.000   & 86351.0  & 156152.1 & 77069.4 \\
225.000   & 56040.8  & 101392.2 & 50044.9 \\
200.000   & 46952.8  & 84963.1  & 41936.9 \\
150.000   & 30481.4  & 55173.6  & 27235.0 \\
75.000    & 10770.7  & 19500.4  & 9628.4 \\
37.500  & 3808.6   & 6893.1   & 3405.3 \\
18.750 & 1347.9   & 2436.7   & 1205.0 \\
9.375 & 477.6    & 861.3    & 426.9 \\
7.400   & 335.4    & 604.0    & 299.6 \\
2.000     & 47.9     & 84.8     & 42.7 \\
\bottomrule
\end{tabular}
\end{center}
%%%%%%%%%%%%%%%%%%%%%%%%%%%%%%%%%
% Don't change these lines
\bsp	% typesetting comment
\label{lastpage}
%%%%%%%%%%%%%%%%%%%%%%%%%%%%%%%%%
\end{document}